\documentclass{aa501}

\usepackage{psfig}
\usepackage{rotating}
\usepackage{amssymb}
\usepackage{graphicx}
\usepackage[innercaption]{sidecap}
\usepackage{natbib}

        \def\cm2{${\rm cm^{-2}}$}              
        \def\cm3{${\rm cm^{-3}}$}              
        \def\mum{${\rm \, \mu m}$}                            

        \newcommand{\Tel}{$T_{\rm e}$}                        
        \newcommand{\Teff}{$T_{\rm eff}$}                     
        \newcommand{\nel}{$n_{\rm e}$}                        
        \newcommand{\rmsnel}{$n_{\rm e,rms}$}                 
        \newcommand{\rgal}{R$_{\rm Gal}$}
        \def\Av{A$\rm _{V}$}
        \def\Ak{A$_{\rm K}$}
        \def\Al{A$_{\lambda}$}
        \newcommand{\Asil}{A$_{\rm 9.7}$} 
        \def\A18{A$_{\rm 18}$} 

        \newcommand{\HI}{\ion{H}{i}}
        \newcommand{\HII}{\ion{H}{ii}}

        \newcommand{\NeII}{[\ion{Ne}{ii}]}
        \newcommand{\NeIII}{[\ion{Ne}{iii}]}
        \newcommand{\OIII}{[\ion{O}{iii}]}
        \newcommand{\NII}{[\ion{N}{ii}]}
        \newcommand{\NIII}{[\ion{N}{iii}]}
        \newcommand{\ArII}{[\ion{Ar}{ii}]}
        \newcommand{\ArIII}{[\ion{Ar}{iii}]}
        \newcommand{\SIV}{[\ion{S}{iv}]}
        \newcommand{\CII}{[\ion{C}{ii}]}
        \newcommand{\OI}{[\ion{O}{i}]}
        \newcommand{\SiII}{[\ion{Si}{ii}]}
        \newcommand{\SIII}{[\ion{S}{iii}]}

        \def\neratio{[\ion{Ne}{iii}]/[\ion{Ne}{ii}]}
        \def\arratio{[\ion{Ar}{iii}]/[\ion{Ar}{ii}]}
        \def\sratio{[\ion{S}{iv}]/[\ion{S}{iii}]}
        \def\nratio{[\ion{N}{iii}]/[\ion{N}{ii}]}

        \def\noionic{N$^{++}$/O$^{++}$}
        \def\nionic{N$^{++}$/N$^{+}$}
        \def\neionic{Ne$^{++}$/Ne$^{+}$}
        \def\arionic{Ar$^{++}$/Ar$^{+}$}
        \def\sionic{S$^{+3}$/S$^{++}$}

\begin{document}

\title{ISO spectroscopy of compact \HII\ regions in the Galaxy
\subtitle{II. Ionization and elemental abundances}
  \thanks{Based on observations with ISO, an ESA project with
    instruments funded by ESA Member States (especially the PI 
    countries: France, Germany, the Netherlands and the United
    Kingdom) and with the participation of ISAS and NASA.} }

        \author{N.L.\,Mart\'{\i}n-Hern\'{a}ndez\inst{1} 
        \and E.\,Peeters\inst{2,1}
        \and C.\,Morisset\inst{3}
        \and A.\,G.\,G.\,M. Tielens\inst{1,2}
        \and P. Cox\inst{4} 
        \and P.\,R. Roelfsema\inst{2}
        \and J.-P\,Baluteau\inst{3}
        \and D.\,Schaerer\inst{5}
        \and J.S.\,Mathis\inst{6}
        \and F.\,Damour\inst{3} 
        \and E.\,Churchwell\inst{6}
        \and M.F.\,Kessler\inst{7}
        }

\offprints{N.L.\,Mart\'{\i}n-Hern\'{a}ndez (leticia@astro.rug.nl)}

\institute{ 
  Kapteyn Institute, P.O. Box 800, 9700 AV Groningen, The Netherlands
  \and SRON, National Institute for Space Reasearch, P.O. Box 800,
  9700 AV Groningen, The Netherlands
  \and Institut d'Astrophysique de Marseille, CNRS \& Univ. de
  Provence, BP 8, F-13376 Marseille Cedex 12, France
  \and Institut d'Astrophysique Spatiale, B\^at. 121, 
  Universit\'e de Paris XI, F-91405 Orsay, France
  \and Laboratoire d'Astrophysique, Observatoire
  Midi-Pyr\'{e}n\'{e}es, 14, Av. E.\,Belin, F-31400 Toulouse, France
  \and Department of Astronomy, 475 North Charter Street, 
  University of Wisconsin, Madison, WI 53706, USA
  \and ISO Data Centre, Astrophysics Division, ESA, Villafranca, Spain
  }

\date{Received date; accepted date}

\titlerunning{Ionization and elemental abundances}
\authorrunning{N.L.\,Mart\'{\i}n-Hern\'{a}ndez et al.}

\abstract{ 
Based on the ISO spectral catalogue of compact \HII\
regions by \cite{peeters:catalogue}, we present a first analysis of the
hydrogen recombination and atomic fine-structure lines originated in
the ionized gas. The sample consists of 34 \HII\ regions located at
galactocentric distances between \rgal =~0 and 15~kpc.  The SWS \HI\
recombination lines between 2 and 8 \mum\ are used to estimate the
extinction law at these wavelengths for 14 \HII\ regions. An
extinction in the K band between 0 and $\sim$ 3 mag. has been  derived.
The fine-structure lines of N, O, Ne, S and Ar are detected in most of
the sources.  Most of these elements are observed in two different
ionization stages probing a range in ionization potential up to 41 eV.
The ISO data, by itself or combined with radio data taken from the
literature, is used to derive the elemental  abundances relative to
hydrogen.  The present data thus allow us to describe for each source
its elemental abundance, its state of ionization and to constrain the
properties of the ionizing star(s).  The main results of this study
are as follows.  The ionization ratios \arionic, \nionic, \sionic\ and
\neionic, which measure the degree of ionization and to first order,
the hardness of the stellar radiation, seem to increase with \rgal.
These ionization ratios correlate well with each other, implying that
the spectral hardening affects equally the full range of ionizing
energies.  A Galactocentric gradient of N/O  ($\Delta \rm log \, N/O =
- 0.056\pm 0.009 \, dex \,kpc^{-1}$) is observed in the sense of a
decreasing  abundance ratio with \rgal\ in agreement with previous
studies.  Abundance gradients for neon and argon are derived of the
form  $\Delta \rm log \, Ne/H = - 0.039\pm 0.007 \, dex \,kpc^{-1}$
and $\Delta \rm log \, Ar/H = -0.045\pm 0.011 \, dex \,kpc^{-1}$.
These elemental gradients could be enlarged by the existing Galactic
\Tel\ gradient. Adopting a \Tel\ gradient of approximately 330
K\,kpc$^{-1}$, the slopes in the Ne/H and Ar/H gradients  become
$-0.06$ and $-0.07$ dex \,kpc$^{-1}$, respectively. Lower limits for
the sulphur and oxygen abundances are derived. Nitrogen abundances are
derived for 16 sources.  
}

\maketitle

\section{Introduction}

The distribution of Galactic elemental abundances is central to study
the chemical evolution of the Milky Way. The relative abundances of
the elements are sensitive to the star formation history, the number
of massive stars, the relative yield of the elements, and the exchange
of matter between the disk and the halo through infall or ejection.
Measurements of elemental abundances throughout the Galactic disk
thus provide vital inputs to model the formation and evolution  of the
Galaxy.

\HII\, regions are prime targets to derive the present-day elemental
abundances. These objects consist of gas which is ionized and heated
by the radiation of massive stars.  When the massive ionizing star is
newly formed, it is still embedded in its natal molecular cloud and
the resulting \HII\, region is a young, bright and compact nebula.
Measuring the elemental abundances of the ionized gas in \HII\,
regions allows us to probe the interstellar medium (ISM) in the
vicinity of massive stars and to trace the present composition of the
ISM in the Galaxy.

Abundance determination methods based on infrared observations
present clear advantages over optical studies because  the infrared
fine-structure lines are not very sensitive to changes in  the
electron temperature and do not suffer from high extinction.  Infrared
measurements can thus trace the elemental abundances in  the central
regions of the Galaxy. The infrared is also the only  wavelength
regime to measure the N$^{++}$ ion, the dominant form  of nitrogen in
highly ionized \HII\ regions.

Previous determinations of the Galactic distribution of elemental
abundances based on the study of \HII~ regions have been performed in
the optical \citep{shaver83,fich91,deharveng00}
and in the infrared, based on IRAS \citep{simpson90}
and Kuiper Airborne Observatory (KAO) observations 
\citep{lester87,rubin88,simpson95a,afflerbach97,rudolph97}.  
These studies resulted in a series of
firm results concerning the variation of elemental abundances across
the Galaxy: the electron temperature (\Tel) of the \HII~ regions
increases with \rgal; the abundance ratio N/O decreases with \rgal;
the abundances of the heavy elements N, O, Ne, S, and Ar decrease with
\rgal. Additional optical studies are based on photospheric emission
lines from B stars 
\citep{fitzsimmons92,smartt97,gummersbach98,rolleston00,smartt01}
and on planetary nebulae \citep{maciel94,maciel99}.  Similar results and 
gradients  have been derived towards other
galaxies \citep[e.g.][]{vila92,zaritsky94}.

\begin{figure*}[ht!]
          \vspace{0.15cm}
          \begin{center}
            \centerline{\psfig{file=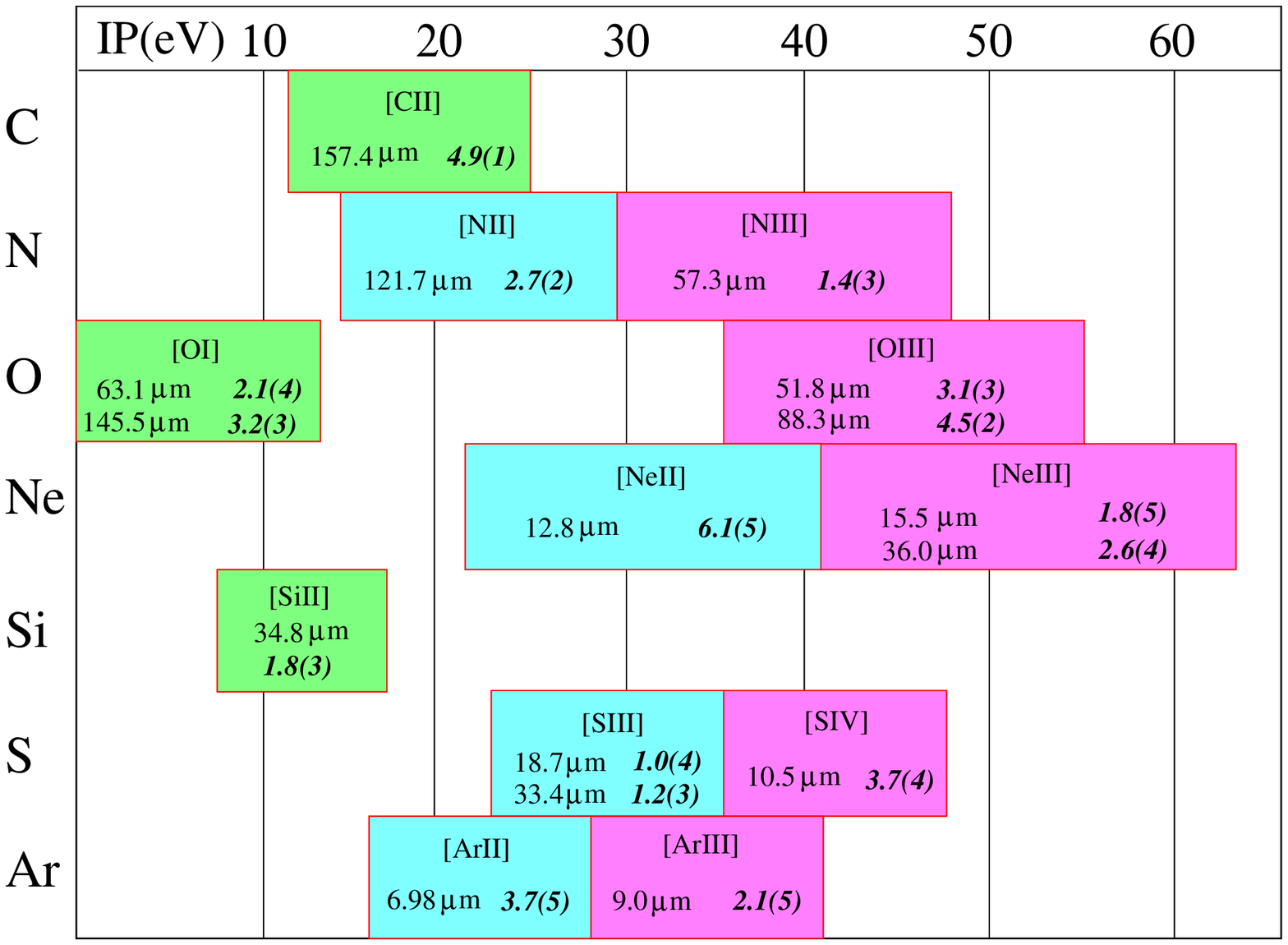,width=14cm}}
          \end{center}
          \vspace{-0.5cm}
          \caption{The fine-structure atomic lines observed in the
            combined ISO SWS/LWS spectra of compact \HII~ regions are 
            shown as a function of the ionization potential. The 
            electron critical
            densities, indicated for every line in italic,  are
            given in units of cm$^{-3}$ and are expressed as 
            $a(b) = a \times 10^b$. The critical densities for the ions
            with ionization potential lower than 13.6~eV (emitted in
            the photodissociation region)
            are taken from \cite{tielens85}. The other
            critical densities were calculated using the latest atomic
            parameters (see Table~\ref{atomicref}).}
          \label{fig:atomic}
        \end{figure*}

The Infrared Space Observatory (ISO) spectral  catalogue of compact
\HII\ regions \citep[][hereafter Paper~I]{peeters:catalogue} presents the
combined Short Wavelength Spectrometer (SWS) and Long Wavelength
Spectrometer (LWS) grating spectra from 2.3 to 196\mum~ for 43
nebulae.  The catalogue tabulates the fluxes for the hydrogen
recombination and atomic fine-structure lines. A detailed explanation
on the error of these line fluxes, which will be used in the present
paper, is given in Sect.~5 of the catalogue.  The spectral coverage gives
access to nearly all the atomic fine-structure lines in the infrared
range. Lines from the elements C, O, N, S, Ne, Si and Ar  are present
in most of the sources.   This paper will concentrate on the lines
emitted by the ionized gas in the \HII\ region.  The \CII, \OI\  and
\SiII\ lines, produced by ions with ionization potentials lower than
13.6 eV, are expected to be mostly emitted in the Photodissociation
Region (PDR) and will be analyzed by Damour et al. (in prep).   For
some of the ions (O$^{++}$, Ne$^{++}$ and S$^{++}$) two lines are
present providing, in principle, an estimate of the electron density.
All the elements except for oxygen (note that O$^{0}$, as mentioned
above, is produced in the PDR) are observed in two ionization stages
(see Fig.\ref{fig:atomic}), which alleviates the problem of applying
ionization correction factors for unseen ions (especially for S, Ne
and Ar).  Finally, the range of ionization potential covered by this
set of data (up to 41~eV) allows us to examine the ionization state of
the \HII\ regions and to constrain the properties of the ionizing
star(s).  From the whole sample, 34 \HII\ regions present enough
emission lines to determine their elemental abundances and ionization
properties.  This subsample covers the Galactic plane from the centre
to a galactocentric distance, \rgal, of 15~kpc. It is thus possible to
investigate trends  of the ionization conditions in \HII\ regions and
of relative  and absolute abundances across a large part of the
Galactic disk.  Atomic parameters are crucial in order to interpret
correctly the  strengths of the different lines.  The latest
transition probabilities and collisional strengths  have been compiled
(cf. Table~\ref{atomicref}) and used throughout this paper.

This paper is structured as follows.  Sect.~\ref{Properties} describes
the kinematic distances and radio properties of the \HII\ regions;
Sect.~\ref{Extinction} discusses the problem of extinction;  a discussion
of the densities of the \HII\ regions is given in  Sect.~\ref{Density};
Sect.~\ref{Methodology} gives  an outline of the methodology used to
derive elemental abundances with a particular emphasis on the
advantages/disadvantages of  the present data;  Sect.~\ref{Ionization}
presents the results for the ionization  state and its variation with
galactocentric distance.  The ionic and elemental abundances, and
their variation with  \rgal, are presented  in Sect.~\ref{Abundances};
Finally, Sect.~\ref{Conclusions} discusses and  summarizes the results of
this paper.

\begin{table}[ht!]
  \caption{References for the transition probabilities A and the
    collisional stengths $\Omega$  used
    in this paper. IP refers to the IRON Project.}
  \label{atomicref}
  \begin{center}
    \leavevmode
    \scriptsize
    \begin{tabular}[h]{lll}
      \hline \\[-5pt]
    \multicolumn{1}{c}{Species} &
    \multicolumn{1}{c}{} &
    \multicolumn{1}{c}{Reference}\\[1pt]
      \hline \\[-5pt]
\ion{N}{II}  & A:       & \cite{galavis97} (IP XXII) \\
             & $\Omega$:& \cite{lennon94} (IP II)\\[2pt]
\ion{N}{III} & A:       & \cite{galavis98a} (IP XXIX)\\
             & $\Omega$:& \cite{blum92} \\[2pt]
\ion{O}{III} & A:       & \cite{galavis97} (IP XXII)\\
             & $\Omega$:& \cite{lennon94} (IP II)\\[2pt]
\ion{Ne}{II} & A:       & \cite{mendoza83a} \\
             & $\Omega$:& \cite{saraph94} (IP IV)\\[2pt]
\ion{Ne}{III}& A:       & \cite{galavis97} (IP XXII)\\
             & $\Omega$:& \cite{butler94} (IP V)\\[2pt]
\ion{S}{III} & A:       & \cite{biemont83} \\
             & $\Omega$:& \cite{tayal99} \\[2pt]
\ion{S}{IV}  & A:       & \cite{mendoza83a} \\
             & $\Omega$:& \cite{saraph99} (IP XXX)\\[2pt]
\ion{Ar}{II} & A:       & \cite{mendoza83a} \\
             & $\Omega$:& \cite{pelan95} (IP IX)\\[2pt]
\ion{Ar}{III}& A:       & \cite{mendoza83b} \\
             & $\Omega$:& \cite{galavis95} (IP X)\\
             &          & \cite{galavis98b} (IP XXXII)\\[2pt] 
             \hline \\
      \end{tabular}
  \end{center}
\vspace{-1cm}
\end{table}

\section{Distances and radio properties}
\label{Properties}

\begin{table*}[!ht]
\caption{Kinematic distances and radio properties of the sample sources.}
\label{table:radio}
\begin{center}
\leavevmode
\footnotesize
\begin{tabular}{lrr@{\,}rccccr@{\,}rc@{\,}cc} \hline \\
\multicolumn{1}{c}{Source}
& \multicolumn{1}{c}{R$_{\rm gal}$}
& \multicolumn{2}{c}{R$_{\sun}$$^{(a)}$}
& \multicolumn{1}{c}{$\lambda$}
& \multicolumn{1}{c}{$S_{\nu}$}
& \multicolumn{1}{c}{Size}
& \multicolumn{1}{c}{$EM$}
& \multicolumn{2}{c}{$n_{\rm e, rms}$$^{(b)}$}
& \multicolumn{2}{c}{$\log N_{\rm Lyc}$$^{(b)}$} 
& \multicolumn{1}{c}{Ref.} \\
\multicolumn{1}{c}{}
& \multicolumn{1}{c}{(kpc)}
& \multicolumn{2}{c}{(kpc)}
& \multicolumn{1}{c}{(cm)}
& \multicolumn{1}{r}{(mJy)}
& \multicolumn{1}{r}{(\arcsec)}
& \multicolumn{1}{r}{(10$^6$ pc cm$^{-6}$)}
& \multicolumn{2}{c}{(10$^3$ cm$^{-3}$)}
& \multicolumn{2}{c}{(s$^{-1}$)} 
& \multicolumn{1}{r}{} \\[2pt] \hline
IR\,01045 & 13.8 &  7.0 &  &  6 &   290 &   3 &  14.5 &  11.9 &  & 47.9 &  
 & 1,2\\
IR\,02219 & 11.0 &  3.3 &  &  6 & 25000 &  35 &   9.2 &   4.0 &  & 49.2 & 
 & 3,4\\
IR\,10589 &  9.5 &  8.0 &  & - & - & - & - & - &   & - &   & -\\
IR\,11143 &  9.7 &  8.8 &  & - & - & - & - & - &   & - &   & -\\
IR\,12063 &  9.3 &  9.5 &  & - & - & - & - & - &   & - &   & -\\
IR\,12073 & 10.1 & 10.8 &  & - & - & - & - & - &   & - &   & -\\
IR\,12331 &  6.9 &  4.5 &  & - & - & - & - & - &  &-  &   & -\\
IR\,15384 &  6.4 &  2.7 & (11.5) & - & - & - & - & - &  &-  &   & -\\
IR\,15502 &  4.6 &  6.0 & (8.4) & - & - & - & - & - &   &-  &   & -\\
IR\,16128 &  5.5 &  3.7 & (11.4) & - & - & - & - & - &  &-  &   & -\\
IR\,17160 &  3.0 &  5.7 & (11.0) &  6 &  1300 &  12 &   4.0 &   3.5 & (2.5) & 
48.4 & (48.9) & 5\\
IR\,17221 &  5.2 &  3.4 & (13.4) & - & - & - & - & - &   & - &   &- \\
IR\,17279 &  3.4 &  5.1 & (11.8) &  6 &   161 &  10 & 0.7 &   1.7 & (1.1) & 
47.3 & (48.1) & 5\\
 Sgr\,C &  0.3 &  8.2 & (8.8) & 21 &  6400 & 120 & 0.2 &   0.2 & (0.2) & 49.3
 & (49.4) & 6\\
IR\,17455 &  0.5 &  8.0 & (9.0) &  6 &  1600 &  14 &   3.7 &   2.6 & (2.4) & 
48.7 & (48.8) & 5,7\\
IR\,17591 &  5.5 &  3.0 & (13.8) &  6 &  1310 &   8 &   9.2 &   8.9 & (4.1) & 
47.8 & (49.1) & 7\\
IR\,18032 &  7.6 &  1.0 & (15.8) &  6 &   640 &  15 &   1.3 &   4.2 & (1.1) & 
46.5 & (48.9) & 7\\
IR\,18116 &  4.3 &  4.5 & (12.0) &  6 &  3860 &  20 &   4.3 &   3.1 & (1.9) & 
48.6 & (49.5) & 7\\
IR\,18162 &  6.6 &  1.9 & (14.7) &  6 &     4 &   5 & 0.07 &   1.3 & (0.5) & 
44.9 & (46.7) & 1\\
IR\,18317 &  4.5 &  4.9 & (10.6) &  6 &  1290 &  13 &   3.4 &   3.3 & (2.3) & 
48.2 & (48.9) & 5\\
IR\,18434 &  4.6 &  5.7 & (9.0) &  6 &  2280 &   7 &  20.9 &  10.4 & (8.3) & 
48.6 & (49.0) & 5\\
IR\,18469 &  4.8 &  5.3 & (9.2) &  6 &   102 &  13 & 0.3 &   0.9 & (0.7) & 
47.2 & (47.7) & 5\\
IR\,18479 &  7.5 &  1.2 & (13.1) &  6 &  2030 &   7 &  18.6 &  21.4 & (6.5) & 
47.2 & (49.3) & 5\\
IR\,18502 &  4.7 &  7.1 &        &  6 &  1090 &   6 &  13.6 &   8.1 & (2.2) & 
48.5 & (50.8) & 7\\
IR\,19207 &  6.1 &  5.7 &        & 21 & 13700 & 140 & 0.3 &   0.3 & (0.1) & 
49.3 & (51.8) & 8\\
IR\,19442 &  7.6 &  2.5 & (5.8) &  6 &    23 &   1 &  10.3 &  29.2 & (19.2) & 
45.9 & (46.6) & 9\\
IR\,19598 &  9.8 &  8.5 &  & 3.6 &  3475 &   6 &  45.6 &  13.6 &  & 49.1 & 
 & 10\\
 DR\,21 &  8.6 &  2.8 &  &  6 & 17500 &  20 &  19.6 &   8.5 &  & 48.9 &  & 11\\
IR\,21190 & 12.7 &  8.9 &  &  6 &   906 &   4 &  25.4 &  12.1 &  & 48.6 & 
 & 12\\
IR\,21270 & 14.8 & 11.3 &  &  6 &   563 & 120 & 0.02 &   0.1 &  & 48.6 &  & 13\\
IR\,21306 & 12.6 &  8.3 &  & - & - & - & - & - &   & - &   & -\\
IR\,22308 & 11.3 &  5.5 &  &  6 &   554 &  90 & 0.03 &   0.1 &  & 48.0 &  & 13\\
IR\,23030 & 11.4 &  5.2 &  &  6 &  1900 &  40 & 0.5 &   0.7 &  & 48.4 &  & 14\\
IR\,23133 & 11.7 &  5.5 &  &  6 &  1048 &  15 &   2.1 &   2.3 &  & 48.2 & 
 & 14\\
\hline
\end{tabular}
\end{center}
{\scriptsize
$^{(a)}$ Value in brackets is the far solar distance.\\
$^{(b)}$ Two different values are derived for sources within the solar
circle because of the near-far solar distance ambiguity. Values in
brackets are calculated using the far solar distance.\\
REFERENCES: 
(1) \cite{mccutcheon91}
(2) \cite{rudolph96}
(3) \cite{roelfsema91}
(4) \cite{tieftrunk97}
(5) \cite{becker94}
(6) \cite{liszt95}
(7) \cite{garay93}
(8) \cite{mehringer94}
(9) \cite{barsony89}
(10) \cite{afflerbach96}
(11) \cite{roelfsema89}
(12) \cite{isaacman84}
(13) \cite{fich93}
(14) \cite{fich86} }
\end{table*}

\subsection{Kinematic distances}

The study of the elemental abundance variation from atomic
fine-structure lines emitted by \HII\ regions relies on proper
determinations of the galactic distances to the \HII\ regions and in
this sense, a detailed literature study has been done in order to  get
the most accurate distances to the program sources.

The determination of the distances to the \HII\ regions of the ISO
sample has been discussed in Paper~I and the reader is referred to
this paper for detailed information. Table~\ref{table:radio} lists in
columns 2 and 3 the adopted distances \rgal\ and $\rm R_\odot$
(including in the latter the near and far solar distances).

\subsection{Radio properties}

Radio continuum measurements allow one to derive various properties of
the \HII\ regions such as the emission measure ($EM$), rms electron
density (\rmsnel) and the Lyman continuum photon flux ($N_{\rm
Lyc}$). These parameters can be derived for spherical sources if the
radio flux density and the source size are known. For non-spherical
sources, the geometrical average of the major and minor  axes can be
taken as a representative size.
  
In the Rayleigh-Jeans limit, the brightness temperature, $T_{\rm b}$,
can be estimated from:

\begin{equation}
{T_{\rm b}}= {{10^{-23} c^2 {S_\nu}} \over {2 \nu^2 k \Omega}}~[{\rm K}],
\label{eq:tb}
\end{equation}

\noindent
where $S_{\nu}$ is the integrated flux density in Jy, 
$\nu$ is the frequency in Hz and $\Omega$
is the source solid angle in sr. The optical depth for an ionized, optically
thin gas at a temperature \Tel\ is given by:

\begin{equation}
  T_{\rm b}= T_{\rm e} (1-e^{-\tau}) \sim T_{\rm e} \tau~.
\label{eq:tau}
\end{equation}

The emission measure can be derived from $\tau$ using: 

\begin{equation}
{EM}= { {\tau} \over 
{8.235\times 10^{-2} a_\nu \left ({T_{\rm e}} \over {\rm K}
  \right)^{-1.35} \left ( \nu \over {\rm GHz} \right )^{-2.1}} }
~[{\rm pc\,cm^{-6}}],
\label{eq:em}
\end{equation}

\noindent
where {\it a}$_\nu$ is a correction factor close to unity 
\citep{mezger67}. 

To estimate the rms electron density we used the formalism explained by 
\cite{wood89}:

\begin{equation}
{n_{\rm e,rms}}= \sqrt { {EM [{\rm pc\,cm^{-6}}]} 
\over {\Delta s[{\rm pc}]}}~[{\rm cm^{-3}}]~,
\label{eq:rmsden}
\end{equation}

\noindent
where $\Delta${\it s} is the path length  of the emitting region 
(taken as the source size). Finally, 
the photon flux of the Lyman continuum can be determined from:

\begin{equation}
{N_{\rm Lyc}} = 1.54\times 10^{55} ({n_{\rm e,rms}})^2 
\left ({{\Delta s} \over {\rm pc}} \right )^3 \alpha_{\rm B}~~[{\rm s^{-1}}]~,
\label{eq:nlyman}
\end{equation}

\noindent
where $\alpha_{\rm B}$=3.28$\times 10^{-13}$ cm$^3$ s$^{-1}$ 
\citep{hummer87} is the case
B recombination coefficient of hydrogen to all levels n$\geq$2 (for 
\Tel=7500\,K).

Flux densities and source sizes were taken from the literature
selecting observations made at frequencies larger than 5 GHz. This
ensures that the emission is optically thin. In addition, care was
taken to select  observations  with a spatial resolution of the order
of a few arcseconds to avoid structures larger than 10\arcsec\ to be
resolved out. Radio properties derived using the above equations are
listed in Table~\ref{table:radio} for those sources with published
observations satisfying the above requirements.

\section{Extinction}
\label{Extinction}

Compact \HII\ regions are heavily embedded in dust and gas and suffer
a dimming of their radiation by dust located either within the ionized
gas or in the neutral foreground material. The term ``extinction''
usually refers to the dimming of starlight, for which scattering from
the beam contributes as much as true absorption because the star is
effectively a point source. For extended objects, the scattering is
less effective in attenuating the radiation because photons can be
scattered into the beam as well as out of it. The effectiveness of
scattering depends upon the geometry of the sources relative to the
dust. In our objects, this is unknown. Hence, we will refer to the
dimming of our objects due to dust absorption as ``extinction''
throughout this paper.

The extinction towards compact \HII\ regions is generally  high with
values of \Av\ ranging from $\sim 10$ to hundreds of magnitudes -
corresponding to extinction of $\sim 1$ to a few 10 mag. at near- and
mid-infrared wavelengths. The extinction in the far-infrared is
generally negligible.  For a number of \HII\ regions, the ISO SWS
spectra show a series of  \HI\ recombination lines which  can be used
to derive the extinction curve  characterizing the dust absorption
towards these sources.  Radio continuum emission can also be used to
derive the expected flux at any \HI\ recombination line to provide an
alternative estimate of the extinction when comparing with the
observed line flux.

\subsection{\HI\ recombination lines method}
\label{Extinction:HI}

        \begin{figure*}[ht!]
          \vspace{0.5cm}
          \begin{center}
            \centerline{\psfig{file=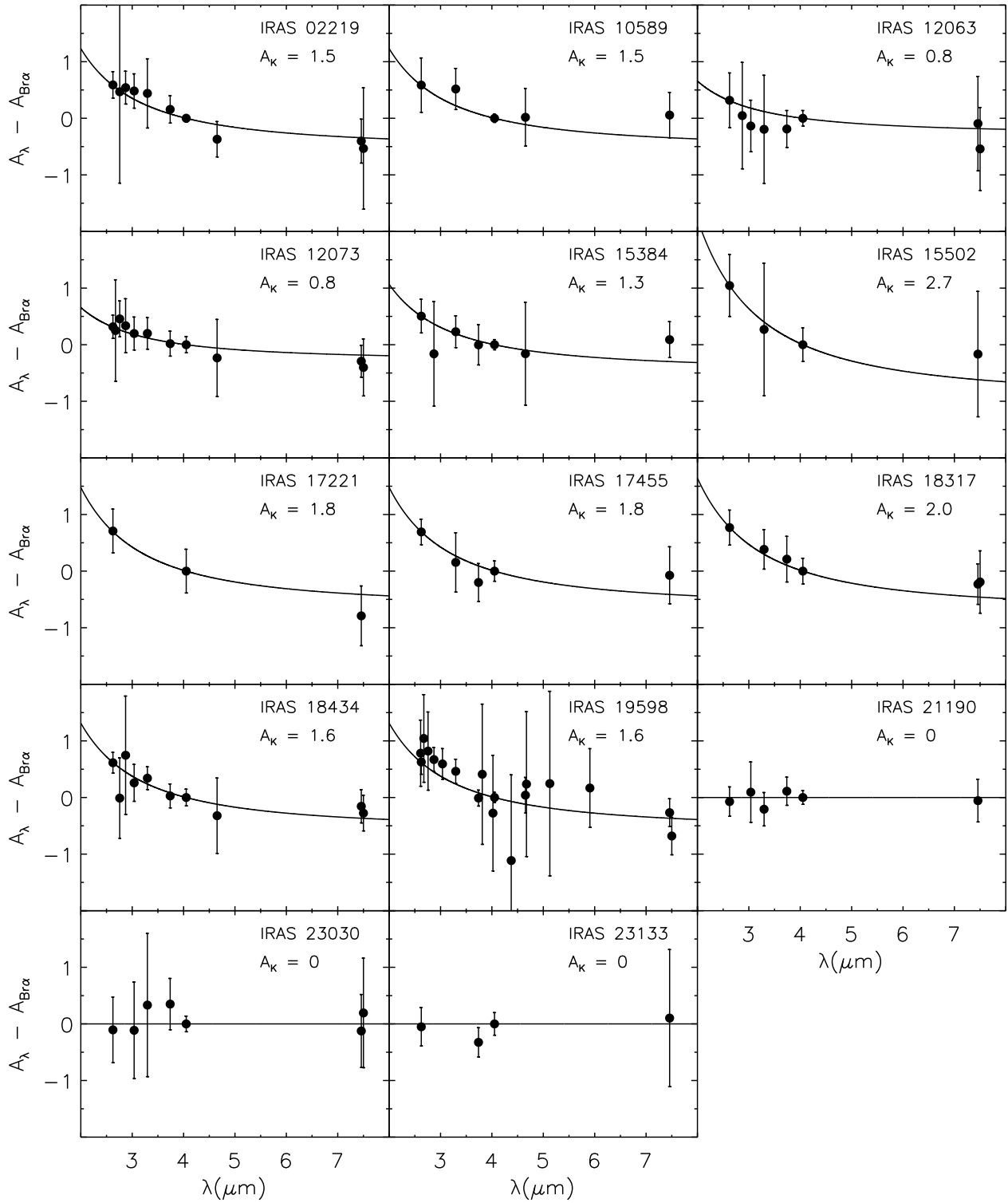,width=18cm}}
          \end{center}
          \caption{Near- and mid-infrared extinction curves as derived
            from the \HI\ recombination lines. Each panel is labeled
            by the name of the \HII\
            region and the derived extinction \Ak. 
            The solid lines show the extinction law 
            (see Sect.~\ref{Extinction:HI}).}
          \label{fig:extinction}
        \end{figure*}

The mid-infrared \HI\ recombination lines  from the Brackett
(n$\rightarrow$4), Pfund (n$\rightarrow$5) and Humpreys
(n$\rightarrow$6) series between 2 and 8~\mum\ can be used to derive
the extinction law at these wavelengths by  comparing the ratio of
observed line strengths to the predictions of recombination theory
assuming a nebular electron density, \nel, and temperature, \Tel.  The
ratio of the hydrogen emission coefficients is not very sensitive to
\nel\ or \Tel. Therefore, we have adopted \nel=1000~\cm3 and
\Tel=7500~K, which are typical values for \HII\ regions.  In the
following, we will consider the standard case B treatment, ie. where
the nebula is optically thick to the Lyman series photons but
optically thin to all other series photons \citep[see, e.g.,][]{osterbrock89}.
The observed line flux, $F({\lambda})$, can then be expressed as:

\begin{equation}
F({\lambda})=F_{\rm o}(\lambda) 10^{-0.4A_{\lambda}}~,
\end{equation}

\noindent
where $F_{\rm o}(\lambda)$ is the non-extincted line flux, given by
case B theory,  and
\Al\ is the extinction at the wavelength $\lambda$. The relative extinction 
to a reference line (e.g. Br$\alpha$ at 4.05\mum) is:

\begin{equation}
{\rm A_{\lambda}-A_{\rm Br\alpha}}= -2.5\left (
{ {\rm log} {F(\lambda) \over F(\rm Br\alpha)}} - 
{\rm log} {F_{\rm o}(\lambda) \over F_{\rm o}(\rm Br\alpha)}\right ).
\end{equation} 

\noindent  
The theoretical ratios are taken from \cite{hummer87}.

Fig.~\ref{fig:extinction} shows the relative extinction ${\rm
A_{\lambda}-A_{\rm Br\alpha}}$ derived for the \HII\ regions which
have at least 3 \HI\ recombination lines (14 sources in total) within
the SWS range.  The line fluxes are taken from Tables~7 \& 8 in
Paper~I.

At infrared wavelengths longward of 1.25~\mum, the extinction law, as
based on JHK photometry, appears to be independent of the line of
sight and consistent with a simple power law ${\rm A_{\lambda}}={\rm
{A_{K}}}(\lambda / 2.2)^{-1.7}$ \citep{mathis90,martin90},
were \Ak\ is the extinction in the K band (2.2~\mum).  For all the 14
sources we could derive an \Ak\ for which this power law satisfies
both the relative extinction at 2.62~\mum\ (Br$\beta$) and 4.05 \mum\
(Br$\alpha$).  The solid lines in Fig.~\ref{fig:extinction} show this
power law for the corresponding \Ak. Note that the curves generally
fit the relative extinction at the other wavelengths nicely.  The
derived extinction values (\Ak) are listed in Table~\ref{table:ext}.

\begin{table}[!ht]
  \caption{Extinction \Ak\ as determined using the \HI\ recombination
    lines method.}
  \label{table:ext}
  \begin{center}
    \leavevmode
    \footnotesize
    \begin{tabular}[h]{lclclc}
      \hline \\[-5pt]
    \multicolumn{1}{c}{Source} &
    \multicolumn{1}{c}{\Ak} &
    \multicolumn{1}{c}{Source} &
    \multicolumn{1}{c}{\Ak} &
    \multicolumn{1}{c}{Source} &
    \multicolumn{1}{c}{\Ak}\\[5pt]
      \hline \\[-5pt]
IR\,02219 & 1.5 & IR\,15502 & 2.7 & IR\,19598 & 1.6 \\
IR\,10589 & 1.5 & IR\,17221 & 1.8 & IR\,21190 & 0.0 \\
IR\,12063 & 0.8 & IR\,17455 & 1.8 & IR\,23030 & 0.0 \\
IR\,12073 & 0.8 & IR\,18317 & 2.0 & IR\,23133 & 0.0 \\
IR\,15384 & 1.3 & IR\,18434 & 1.6 & \\[5pt]
\hline \\
      \end{tabular}
  \end{center}
\vspace{-0.5cm}
\end{table}

\begin{table}[!ht]
  \caption{Estimates of the extinctions \Al/\Ak\ to apply to the near-
    and mid-infrared lines using the ``standard'' extinction law
    described in the text.}
  \label{table:standard}
  \begin{center}
    \leavevmode
    \footnotesize
    \begin{tabular}[h]{lcclcc}
      \hline \\[-5pt]
    \multicolumn{1}{c}{Line} &
    \multicolumn{1}{c}{$\lambda$(\mum)} &
    \multicolumn{1}{c}{\Al/\Ak} &
    \multicolumn{1}{c}{Line} &
    \multicolumn{1}{c}{$\lambda$(\mum)} &
    \multicolumn{1}{c}{\Al/\Ak}\\[5pt]
      \hline \\[-5pt]
Br$\beta$   &2.6   & 0.741 & \NeII\      &12.8  & 0.165 \\
Br$\alpha$  &4.0   & 0.354 & \NeIII\     &15.5  & 0.156 \\ 
\ArII\      &7.0   & 0.140 & \SIII\      &18.7  & 0.217 \\
\ArIII\     &9.0   & 0.429 & \SIII\      &33.5  & 0.069 \\ 
\SIV\       &10.5  & 0.429 & \NeIII\     &36.0  & 0.060 \\[5pt] \hline \\
      \end{tabular}
  \end{center}
\vspace{-0.6cm}
\end{table}


From the computed \Ak\ one can derive the extinction at other
wavelengths  using the results tabulated in \cite{mathis90}, namely
$A_{9.7}/A_{\rm K}=0.544$ for the 9.7~\mum~ silicate feature and
$A_{18}/A_{\rm K}=0.217$ for the 18~\mum~ silicate, in combination
with the astronomical silicate profile of \cite{draine85} and a
$\lambda^{-2}$ power law for $\lambda>20$\,\mum. Using this
``standard'' extinction  law, the extinction  \Al/\Ak\  to apply to
the near- and mid-infrared  line fluxes is shown in
Table~\ref{table:standard}.  In contrast to the extinction between 2
and 7\,\mum, the extinction due to the silicate absorption at 9.7 and
18\,\mum\ depends on the line-of-sight \citep{draine89} and still both
the shape and strength of these features remain controversial.
Therefore, the  ``standard'' extinctions for wavelengths longer than 7
\mum\ given in Table~\ref{table:standard} are merely illustrative and
may not be applicable to all the \HII\ regions.

Because of the lack of information on the extinction for most of the
sample sources (extinction could only be derived for 14 sources out of
34) and the difficulties explained above in correcting  the lines
beyond 7 \mum, {\it the line fluxes will not be  corrected for
extinction} in the analysis presented in this paper and instead the
effect of the extinction will be indicated.

\subsection{Radio continuum method}
\label{Extinction:radio}

        \begin{figure}[!ht]
          \vspace{-0.55cm}
          \begin{center}
            \centerline{\psfig{file=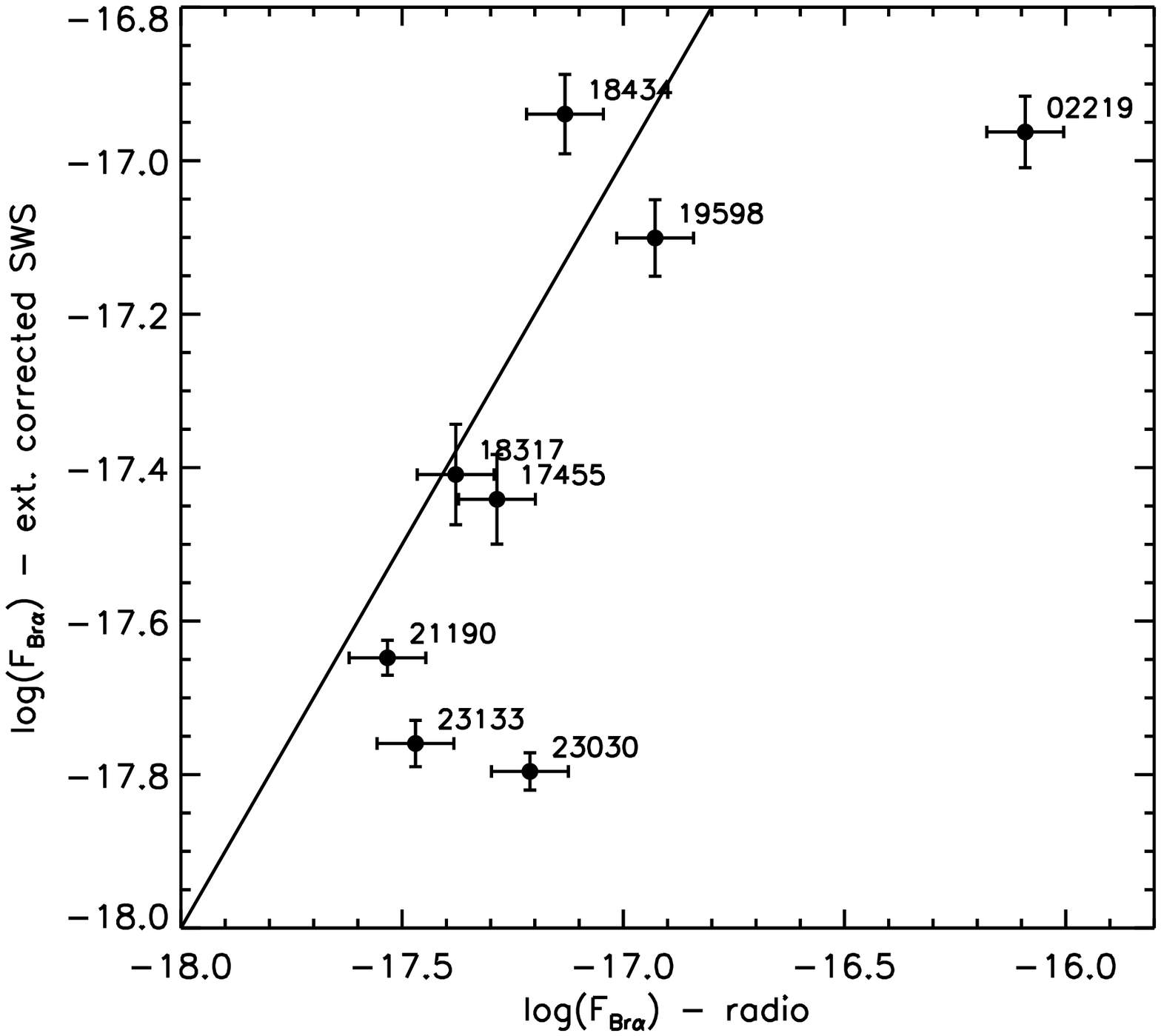,width=8.5cm}}
          \end{center}
          \vspace{-1.5cm}
          \caption{Comparison between the Br$\alpha$ flux predicted 
            by radio observations and the extinction-corrected
            Br$\alpha$ flux observed by SWS. The solid line represents
            the one-to-one relation.}
          \label{fig:braradio}
        \end{figure}

The intrinsic flux of any \HI~ recombination line   can be determined
if both \Tel~ and the emission measure, $EM$,  in the nebula are
known. Radio continuum observations can provide this information (see
Sect.~\ref{Properties}).  The intrinsic flux of Br$\alpha$ can be
determined using:

\begin{equation}
F_{\rm Br\alpha}= \epsilon_{\rm Br\alpha} 
{EM}  \left ( { {n_p} \over {n_e} } \right ){\Omega \over 4\pi}\,,
\label{eq:flux_general}
\end{equation} 

\noindent 
where $\epsilon _{\rm Br\alpha}$ is the emissivity of ${\rm Br}\alpha$
and $\Omega$ is the solid angle of the source. 
Using equations~\ref{eq:tb}, \ref{eq:tau}, \ref{eq:em} and 
\ref{eq:flux_general}, the ratio 
of protons to electrons $n_p/n_e \approx 0.90$, \Tel=7500\,K and
$\epsilon _{\rm Br\alpha}$=1.39$\times 10^{-26}$ erg~s$^{-1}$~cm$^{3}$
\citep{hummer87}, we obtain:

\begin{equation}
F_{\rm Br\alpha}=2.76\times10^{-18}\left({ \nu \over {\rm GHz}} \right )^{0.1} 
\left ( {{\rm S_\nu}\over {\rm Jy}} \right )~~[{\rm W\,cm^{-2}}]\,,
\label{eq:flux_bralpha}
\end{equation}

\noindent
where the flux densities, S$_{\nu}$, from Table~\ref{table:radio} can
be used.  Fig.~\ref{fig:braradio} compares these predicted fluxes for
Br$\alpha$ with the fluxes observed by SWS corrected using  the
extinction derived in ~Sect.~\ref{Extinction:HI}.  The radio flux
densities and A$_{\rm Br\alpha}$ are considered to have a nominal 20\%
uncertainty.  In general, the radio Br$\alpha$ fluxes are larger than
the  extinction-corrected SWS fluxes, especially  for IR\,02219 and
IR\,23030. These two sources are much more extended  in radio than the
SWS beam  (see Table~\ref{table:radio}). On the other hand, the
predicted radio line flux for IR\,18434 is slightly smaller. This
source lies in a region of extended emission of up to a few
arc-minutes \citep{kim01} which is observed by SWS and it is not
included in the quoted radio flux density. This is probably the reason
why the radio prediction leads to a lower line flux.

The application of predictions from radio continuum observations to
our ISO data depends on the source geometry at both radio and infrared
wavelengths and on the knowledge of the radio emission corresponding
to the infrared emission observed by ISO.  We will return to this
issue in Sect.~\ref{Density}.
 
\subsection{Comments on individual sources}

In the following, we comment on the  extinction determination in
individual \HII\ regions from which previous studies are available. 

\vspace{-0.3cm}
\paragraph{IRAS\,02219$+$6125 (W3\,A)}

It is a very bright \HII\ region located in a
large 4' by 3' \citep{dickel80} complex of \HII\ regions, infrared
sources and molecular clouds.  \cite{hayward89} studied the
extinction to W3\,A by comparing the 5 GHz map by \cite{harris76} 
with their infrared (H and K) observations. They
found that the extinction varies from \Ak\ $\sim$ 1.0 to 1.7, in
agreement with our computed extinction (\Ak=1.5). The 9.7 \mum\
extinction was studied by \cite{hackwell78}, who found \Asil\ to
vary from $\sim$2.0 at the position of W3\,A to $\sim$4.0 at the
perimeter of the W3 complex.  This extinction due to the 9.7 \mum\
silicate absorption is much larger than the  \Asil=0.8 derived from
our \Ak\ when the ``standard'' extinction law  described in
Sect.~\ref{Extinction:HI} is used.

\vspace{-0.3cm}
\paragraph{IRAS\,12073$-$6233}

Near-infrared slit observations by \cite{armand96} mapped the
variation of several \HI\ recombination lines across
IRAS\,12073$-$6233 and found \Ak\ to vary from 1.2 to 1.7. We derive
an extinction of \Ak=0.8 mag.   The discrepancy in $\sim 0.4 -
0.9$~mag. can be caused by the fact that we have observed the
integrated \HI~ emission in a $14\arcsec \times 20\arcsec$ beam, which
may lead to a lower hydrogen density column because of the
contribution of extended emission around the nebula and thus, to the
lower extinction we obtain.

\vspace{-0.3cm}
\paragraph{IRAS\,18434$-$0242 (G29.96$-$0.02)}

It is one of the best studied ultracompact \HII~
regions and classified as cometary \citep{wood89}.   By
comparing images of the 2.17~\mum~ Br$\gamma$ recombination line and
2~cm radio continuum emission \citep{fey95}, \cite{watson97a}
derived a map of the extinction towards this \HII~ region. They found
that the apparent extinction to the nebula is not uniform, with \Ak\
varying from 2.2 to 2.6 mag. They give a mean value \Ak=2.14$\pm$0.25.
\cite{pratap99} estimated \Ak=2.2$\pm$0.1 towards the ionizing
star based on its H-K colour. As discussed above in the case of
IRAS\,12073$-$6233, the contribution of the extended envelope around
the compact \HII\ region \citep{kim01} to the SWS aperture may
cause ISO to observe a lower hydrogen column density than the ones
observed by the high $\sim$1\arcsec\ resolution observations of 
\cite{watson97a} and \cite{pratap99}. This could explain our lower
\Ak=1.6 mag.

\vspace{-0.3cm}
\paragraph{IRAS\,19598$+$3324 (K3-50\,A)}

It is the only \HII\ region in the sample
for which the SWS spectrum was also taken at speed 4, yielding a
resolving power of $\sim$~1500 (see Paper~I). In total, 20   \HI\
recombination lines were detected  (Tables~7 \& 8 in Paper~I)
allowing us to derive a better sampled extinction curve in K3-50\,A than
for the other nebulae of the catalogue, especially in the region from
2 to  5~\mum\  (Fig.~\ref{fig:extinction}).   Note that the points
around 3\mum\ seem to be slightly above the standard  extinction
curve, which could be due to the 3\mum\ O--H stretching  mode seen in
the spectrum (see Paper~I).
The extinction to K3-50\,A has been studied in detail by \cite{howard96} 
using both high-resolution near-infrared and radio images. They
found variations in the extinction across the nebula.  They reported
an extinction at 4.05\mum\ (Br$\alpha$) varying from 0 to 3 mag., with
an averaged value of 1.6 in the central 2.7\arcsec\ region. This is
the same value we get from the  \HI\ recombination lines analysis.

\section{Density}
\label{Density}

The electron density (\nel) can be determined from the ratio of two
atomic fine-structure lines of the same ionic species when they are
emitted from levels with nearly the same excitation energy 
\citep[e.g.][]{rubin94}. This ratio is sensitive to gas  densities 
approximately
in between the critical densities of each line  (see
Fig.~\ref{fig:atomic}). A good indicator of gas with  densities lower
than a few thousands particles/cm$^3$ is  the ratio of the \OIII\
lines at 52 and 88~$\mu$m.  Higher density gas can be probed using
line ratios with higher critical densities such as \SIII\ 33/19~\mum\
and \NeIII\ 36/15~\mum.

\subsection{[O\,III] diagnostic}
\label{denOIII}

The densities derived from the \OIII\ line ratio for the ISO sample of
\HII\ regions are in the range from $\sim$~100 to  3000~cm$^{-3}$
(see Table~\ref{table:o3den}). No dependence of  the density on
galactocentric distance is found (see Fig.~\ref{fig:o3denrgal}), in
agreement with previous studies  \citep[e.g.][]{simpson95a,afflerbach97}.

The derived \OIII\ densities, \nel(\OIII), are  compared to the rms
densities, \rmsnel, calculated from the radio continuum observations
(see Table~\ref{table:radio}) in Fig.~\ref{fig:rmsden}. In general,
the rms densities are larger than the \OIII\ densities, except for
three sources (IR\,19207, IR\,21270 and IR\,22308), which are more
extended (their radio sizes are 140, 120 and 90\arcsec, respectively)
than the LWS aperture.

\begin{table}[!ht]
\caption{Electron densities derived from the [O\,III] 88/52 $\mu$m line ratio.}
\label{table:o3den}
\begin{center}
\leavevmode
\footnotesize
\begin{tabular}{lr@{\hspace{1pt}}r@{\hspace{3pt}}llr@{\hspace{1pt}}r@{\hspace{3pt}}l} \hline \\[1pt]
\multicolumn{1}{c}{Source}
& \multicolumn{3}{c}{$n_{\rm e}$([O\,III])}
& \multicolumn{1}{c}{Source}
& \multicolumn{3}{c}{$n_{\rm e}$([O\,III])} \\
\multicolumn{1}{c}{}
& \multicolumn{3}{c}{(cm$^{-3}$)}
&\multicolumn{1}{c}{}
& \multicolumn{3}{c}{(cm$^{-3}$)} \\[2pt] \hline \\[1pt]
IR\,02219 & &    2834 & $^{+    1225} _{-     680}$ & IR\,18116 & &     753 & 
$^{+     224} _{-     152}$ \\[2pt]
IR\,10589 & &     645 & $^{+     194} _{-     132}$ & IR\,18317 & &    1543 & 
$^{+     738} _{-     397}$ \\[2pt]
IR\,11143 & &     290 & $^{+     103} _{-      71}$ & IR\,18434 & &     817 & 
$^{+     260} _{-     171}$ \\[2pt]
IR\,12063 & &    1335 & $^{+     454} _{-     284}$ & IR\,18469 & &     146 & 
$^{+      65} _{-      46}$ \\[2pt]
IR\,12073 & &     962 & $^{+     295} _{-     195}$ & IR\,18479 & &     836 & 
$^{+     268} _{-     175}$ \\[2pt]
IR\,12331 & &     623 & $^{+     181} _{-     125}$ & IR\,18502 & &    1181 & 
$^{+     411} _{-     257}$ \\[2pt]
IR\,15384 & &    1224 & $^{+     390} _{-     251}$ & IR\,19207 & &     485 & 
$^{+     152} _{-     104}$ \\[2pt]
IR\,15502 & &     452 & $^{+     178} _{-     114}$ & IR\,19598 & &     484 & 
$^{+     154} _{-     105}$ \\[2pt]
IR\,16128 & &     891 & $^{+     271} _{-     180}$ &  DR\,21 & &     320 & 
$^{+     139} _{-      90}$ \\[2pt]
IR\,17160 & &     476 & $^{+     144} _{-     100}$ & IR\,21190 & &    2882 & 
$^{+    1987} _{-     871}$ \\[2pt]
IR\,17221 & &     594 & $^{+     182} _{-     124}$ & IR\,21270 & &     186 & 
$^{+      74} _{-      52}$ \\[2pt]
IR\,17279 & &     127 & $^{+      61} _{-      44}$ & IR\,21306 & &     214 & 
$^{+     109} _{-      70}$ \\[2pt]
 Sgr\,C & &     107 & $^{+      66} _{-      45}$ & IR\,22308 & &     768 & 
$^{+     366} _{-     207}$ \\[2pt]
IR\,17455 & &     548 & $^{+     181} _{-     121}$ & IR\,23030 & &     806 & 
$^{+     246} _{-     164}$ \\[2pt]
IR\,17591 & &     462 & $^{+     143} _{-      99}$ & IR\,23133 & $<$ &
     478 & \\[2pt]
IR\,18032 & &     512 & $^{+     274} _{-     154}$ \\[2pt]
\\ \hline
\end{tabular}
\end{center}
\end{table}


If the radio and ISO sample the same gas and there is no collisional
de-excitation, \nel(\OIII) reflects the mean of the local density
and must always be of the same order or greater than \rmsnel,
which is derived assuming that the emission is spread uniformly along
the line of sight. The ratio of both densities is the filling factor,
which has been found to be typically of the order of $\sim$ 0.1 for
many \HII\ regions \citep{copetti00}. The sources with \rmsnel $<$
\nel(\OIII) likely contain high density clumps embedded in a hot, low
density gas from stellar winds.  However, Fig.~\ref{fig:rmsden} shows
that \nel(\OIII) is much too low for most of the sources.  At the
densities implied by the radio continuum observations, the \OIII\ line
at 88 \mum\ (with n$_{\rm crit}= 450~{\rm cm^{-3}}$) is collisionally
de-excited in the ultracompact core of the \HII\ region. Considering
that the real densities of these cores are probably even higher than
our derived rms densities, even the \OIII\ line at 52 \mum\ (n$_{\rm
crit}=3.1\times10^3~{\rm cm^{-3}}$) could also well be  collisionaly
de-excited.  Therefore, the density derived from the \OIII\ lines is
likely representative of a dilute shell surrounding the ultracompact
core. This low density gas at high excitation (note that O$^{++}$ is
produced by ionizing photons with $h\nu > 35$ eV) could be produced by
leakage of radiation from the core.  Recent radio observations 
\citep{kurtz99,kim01} have shown that the presence of
physically related envelopes around  ultracompact \HII\ regions is the
usual case rather than the exception.
  
\subsection{[S\,III] and [Ne\,III] diagnostics} 

The derivation of electron densities using the \SIII\ and \NeIII\ line
ratios from the ISO data yielded inconsistent results.
Fig.~\ref{fig:otherden} compares the observed line ratios of these
ions with  their expected values  for uniform electron densities
between 10$^2$ and 10$^6$ \cm3, plotted as a solid line. The assumed
\Tel\ of $7500$~K has almost no effect on the plot. Meaningful \SIII\
and \NeIII\ line ratios should lie below their expected values at low
densities, but we see that almost  all points for the \NeIII\ line
ratio, and many for \SIII,  lie above these limits by up to a factor
of 3.  This problem has also been recognized for the \NeIII\ line
ratio in some planetary nebula \citep{rubin00}. However, in that
case the observed ratio is only 10\% larger than what theory predicts
for the low density limit.  Possible causes of this discrepancy
between observations and theory are:

       \begin{figure}[!ht]
         \vspace{-0.55cm}
        \begin{center}
           \centerline{\psfig{file=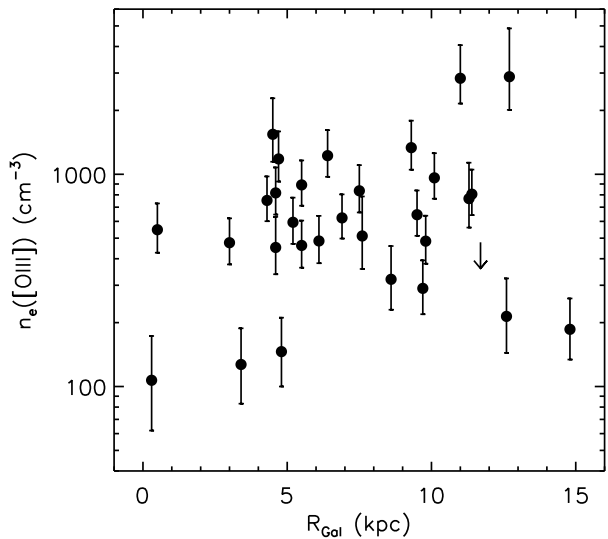,width=8.5cm}}
         \end{center}
         \vspace{-1.5cm}
         \caption{Electron densities of the compact \HII\ regions derived from
           the \OIII\ 88/52~\mum\ line ratio plotted against
           \rgal. The arrow indicates an upper limit.}
         \label{fig:o3denrgal}
       \end{figure}

       \begin{figure}[!ht]
         \vspace{-0.55cm}
         \begin{center}
           \centerline{\psfig{file=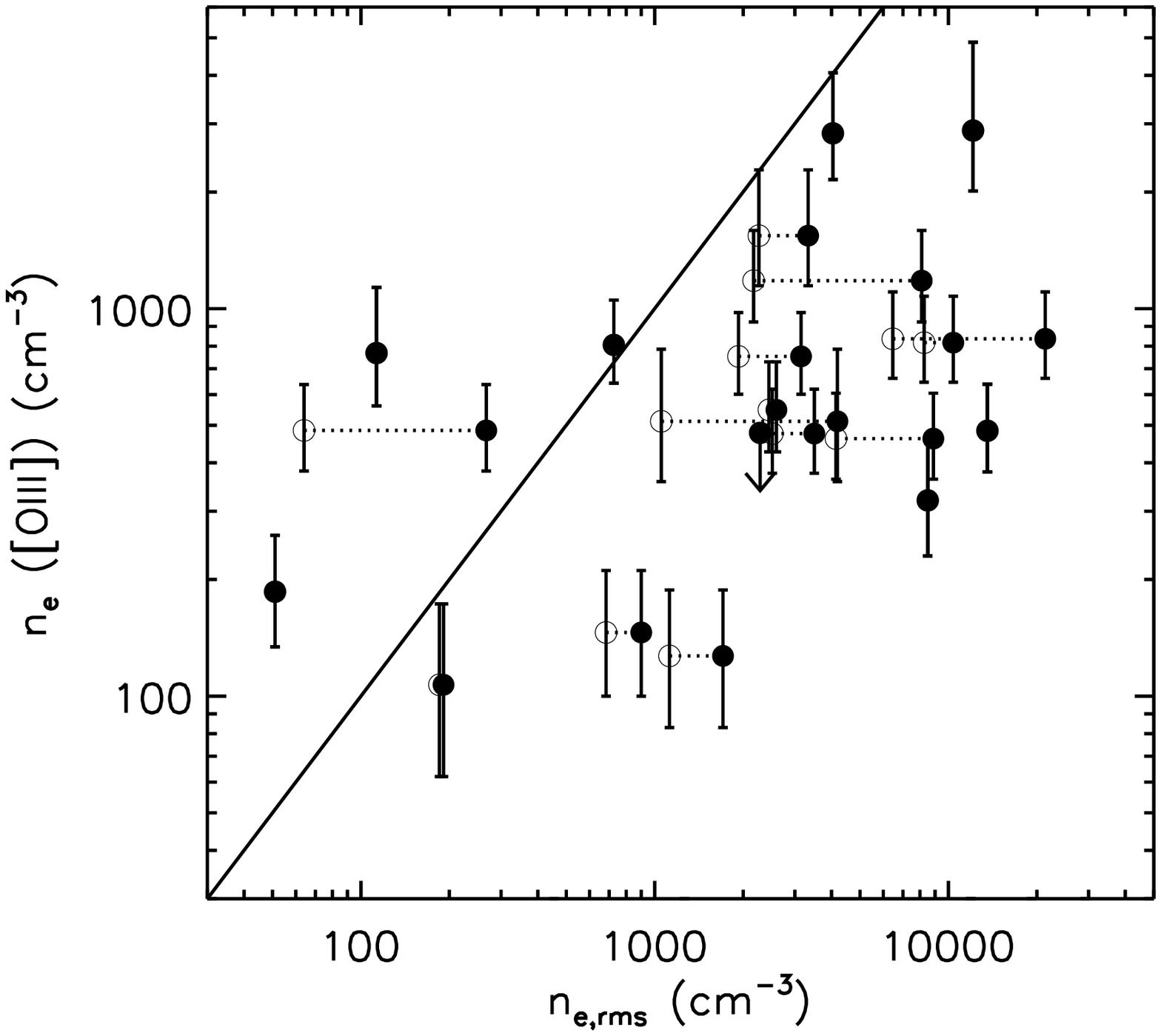,width=8.5cm}}
         \end{center}
         \vspace{-1.5cm}
         \caption{Comparison between the electron densities of the 
           compact \HII\ regions derived from
           the \OIII\ 88/52~\mum\ line ratio,  \nel(\OIII), and the radio
           continuum, \rmsnel. 
           The dotted lines join the \rmsnel\ as derived using
           both the near and far solar distances.
           The solid line represents the one-to-one relationship. The
           fact that \nel(\OIII) is generally lower than 
           \rmsnel\ shows that the \OIII\ lines are 
           collisionally de-excited in the central regions of the
           objects, so that ISO LWS observes only a surrounding shell
           of low density gas.}
         \label{fig:rmsden}
       \end{figure}

\vspace{-0.4cm}
\paragraph{Aperture}

The lines involved in both \SIII\ and \NeIII\ ratios  have been
observed in different apertures ($14\arcsec \times 27\arcsec$ and
$20\arcsec \times 33\arcsec$, respectively for the short and long
wavelength lines).  The black, large arrows in the bottom right corner
of Fig.~\ref{fig:otherden} indicate the shift in both line ratios that
might be caused by aperture effect if the emitting region is larger
than the SWS band 4 ($20\arcsec \times 33\arcsec$) and has a uniform
brightness.  Aperture  effects may have affected the sources which
cluster around a \SIII\ ratio of unity.  Actually, this correction
would bring these sources within the  expected \SIII\ and \NeIII\
ranges at densities of about $10^3-10^4$ \cm3, the rms density range.
However, this aperture correction is not large enough to bring the
most-offending  sources (IR\,16128, IR\,17160, IR\,17279, IR\,17591,
IR\,18469 and IR\,19207), with a \SIII\ ratio $\sim 3$,  into the
theoretical limits. Instead, these  sources are undoubtedly off
because only part of these sources were  included in the SWS
beam. Specifically, recent 6 cm ATCA maps of IR\,16128,  IR\,17160 and
IR\,17279 (Mart\'{\i}n-Hern\'{a}ndez et al. 2002, in prep) show  that
SWS was pointing at the edge of these sources and more of the source
was included in the larger aperture of the longer wavelength lines
than in that of the shorter wavelength lines. Likewise, radio maps of
the sources IR\,17591 \citep{garay93}, IR\,18469 \citep{kurtz94}  
and IR\,19207 \citep{mehringer94} reveal complex structure
-\,core halo or multiple components\,- part of which was not included
in the short  wavelength, smaller aperture.

\vspace{-0.4cm}
\paragraph{Calibration}

The lines at the longest wavelengths, \SIII~33 and \NeIII~36 \mum, are
located in the SWS band 4. Band 4 detectors suffer strongly from
memory effects, are very sensitive to cosmic particle hits and their
responsivity curve is still unsatisfactory and based on pre-flight
ground based tests (see Paper~I). Therefore, these lines are affected
by large calibration errors (an official $1 \sigma$ 25\% calibration
error is quoted in Paper~I and  used in Fig.~\ref{fig:otherden}).
However, this calibration uncertainty is not enough to explain the
discrepancy  between the observed and the predicted line ratios.

\vspace{-0.4cm}
\paragraph{Atomic parameters}

The most recent theoretical calculations \citep{mclaughlin00} of
the collisional strengths for the \NeIII\ levels differ from
earlier  studies \citep{butler94} by only a few
percent. Hence, we consider  it unlikely that the observed factor 2-3
discrepancy in the \NeIII\ line  ratios of \HII\ regions results from
errors in the theoretical collisional  strengths.

\vspace{-0.4cm}
\paragraph{Extinction}

The lines involved in both \SIII\ and \NeIII\ ratios suffer also from
differential extinction. The color excesses between the \SIII\ lines
and the \NeIII\ lines  are $E\simeq 0.15$\,\Ak\ and $E\simeq
0.10$\,\Ak\ (cf. Table~\ref{table:standard}), respectively.   For the
nominal \Ak$\sim 2$ mag. derived from the \HI\ recombination lines
(cf. Table~\ref{table:ext}), these extinction  corrections (indicated
in Fig.~\ref{fig:otherden} by the grey arrows in the bottom right
corner) are, however, quite small.

       \begin{figure}[!ht]
         \vspace{-0.55cm}
       \begin{center}
       \centerline{\psfig{file=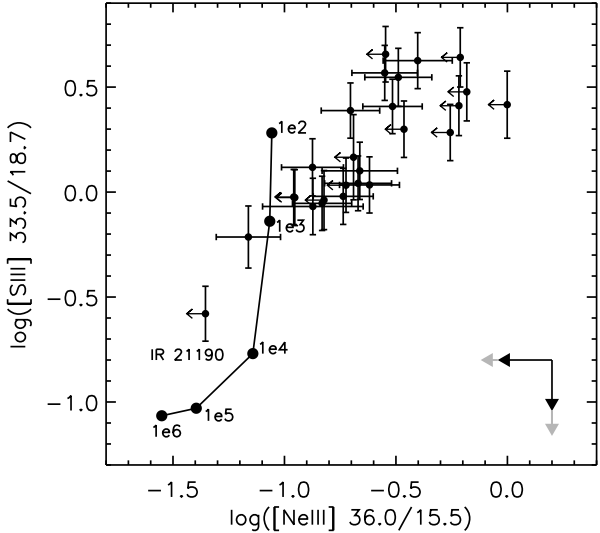,width=8.5cm}}
       \end{center}
         \vspace{-1.5cm}
       \caption{Comparison between the observed line ratios \SIII\
         33/19 \mum\ and \NeIII\ 36/15 \mum. Upper limits for the line
         ratios are indicated.  The solid line represents the
         theoretical values of the [S\,III] and [Ne\,III] line ratios
         for the electron densities 10$^2$, 10$^3$, 10$^4$, 10$^5$ and
         10$^6$ \cm3.  The large, black arrows in the bottom right
         corner indicate the  shift ($\sim 0.23$ in logarithmic scale)
         due to beam correction for an extended, uniform source. They
         are followed by short, grey arrows which indicate the
         extinction correction due to an \Ak=2 mag.  The source
         IR\,21190 is indicated (see text).}
       \label{fig:otherden}
       \end{figure}

\vspace{-0.4cm}
\paragraph{ }

We conclude that the present \SIII\ and \NeIII\ line ratio
observations are unreliable as density indicators largely due to
aperture  differences.  The sources above mentioned (IR\,16128,
IR\,17160, IR\,17279, IR\,17591,  IR\,18469 and IR\,19207), 
together with IR\,21270 and IR\,21306, which, although not present
in Fig.~\ref{fig:otherden} are known to be misspointed (see
Paper~I), will be
excluded from the analysis involving any SWS line. Moreover, the
\SIII\ 33 and \NeIII\ 36 \mum\ lines will not be used in the data
analysis.

The object IR\,21190, identified in Fig.~\ref{fig:otherden},  is of
special interest.  It is a very compact \HII\ region (and thus
point-like for the SWS; see its spectrum in Fig.~8, Paper~I, which
does not present jumps between the different apertures) and has an
\Ak=0 (see  Table~\ref{table:ext}). Thus, it should not be affected by
aperture changes or extinction. Indeed, the \SIII\ and \NeIII\ ratios
for this source are well inside the theoretical limits. However, the
densities derived from these two indicators  (\nel\ $<10^4$ \cm3  and
$>10^5$ \cm3, respectively for the \SIII\ and \NeIII\ ratios) do not
agree. IRAS 21190 presents a  complex density morphology (Peeters et
al.  2002, in prep) and it is  therefore not surprising that the
tracer of lower density and lower ionization state material yields a
lower density than the tracer of higher density and higher ionization
state material.

\section{Abundance determinations}
\label{Methodology}

\subsection{Methodology}

Ionic abundances can be determined from the measured strengths of the
lines when a specific model for the structure of the \HII\ region is
assumed. The simplest model is to assume that the nebula is
homogeneous with constant \Tel\ and \nel. If one also assumes that all
the line photons emitted in the nebula escape without absorption and
therefore without causing further upward transitions, the calculation
of the ionic abundances is straightforward \citep[e.g.][]{rubin88}.

Considering two ions $X^{+i}$ and $Y^{+j}$, the ratio of their
ionic abundances is given by:

\begin{equation}
{ X^{+i} \over Y^{+j} } ={ { F_{X^{+i}} / F_{Y^{+j}} } \over  {
\epsilon_{X^{+i}} / \epsilon_{Y^{+j}} } }~,
\label{eq:ionicab}
\end{equation}

\noindent
where $F_{X^{+i}}$ and $F_{Y^{+j}}$ are the fluxes corrected from
extinction corresponding to any line produced by the ions $X^{+i}$ and
$Y^{+j}$, and $\epsilon_{X^{+i}}$ and $\epsilon_{Y^{+j}}$ are their
respective emission coefficients.  This equation assumes that the
volume occupied by both  ionic gases and the solid angles included in
the fluxes are the same.  Because the beam sizes of ISO differ
significantly with wavelength (see Paper~I), the
distribution of the emission over the beam becomes important in
determining abundances for some of the elements.

\begin{figure}[!ht]
         \vspace{-0.55cm}
         \begin{center}
           \centerline{\psfig{file=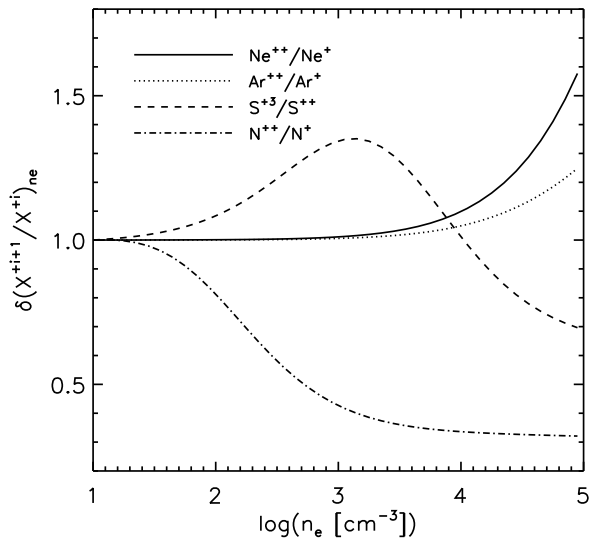,width=8.5cm}}
         \end{center}
         \vspace{-1.5cm}
         \caption{The \nel\ correction factor
           $\delta(X^{+i+1}/X^{+i})_{\rm n_e}$ needed to apply to the 
           ionic ratio $X^{+i+1}/X^{+i}$ when the line
           emission coefficients are evaluated in the low density
           limit is plotted as a function of \nel.}
         \label{fig:ioniz_density}
       \end{figure}

The emission coefficients depend on \Tel, \nel\ and the relevant
atomic parameters (transition probabilities, $A$, and collisional
strengths, $\Omega$). They are computed using a 5-level atom (or a
2-level atom for ions with a $^2P$ ground term) in statistical
equilibrium \citep{kafatos80}. The references for the atomic
parameters used in this paper are given in Table~\ref{atomicref}.
Fortunately, the emissivities of the fine-structure lines show only a
very slight dependence on  \Tel\ because these lines are emitted from
levels with excitation energies much lower than the mean colliding
electron energy. Thus, we will use an electron  temperature of 7500~K,
typical for galactic \HII\ regions 
\citep{shaver83,afflerbach96,afflerbach97}, 
to evaluate these emissivities.
However, the dependence of the fine-structure emission coefficients on
\nel\ is particularly important when the density of the emitting
region is comparable to the critical density of the levels involved.
It is illustrative to consider the \nel\ correction factor,
$\delta(X^{+i+1}/X^{+i})_{n_{\rm e}}$, needed to apply to the
abundance ratio of  sequential ionic ionization states (\neionic,
calculated from \NeII\ 12.8 and \NeIII\ 15.5 \mum, \arionic, from
\ArII\ 7.0 and \ArIII\ 9.0 \mum, \sionic\, from \SIII\ 18.7 and \SIV\
10.5 \mum,  and \nionic, from \NII\ 122 and \NIII\ 57 \mum) when the
respective  line emissivities are evaluated in the low density limit
(cf. Fig.~\ref{fig:ioniz_density}). This correction factor is,
basically, the quotient between the ratio of the respective
emissivities evaluated at a density \nel\ and in the low density
limit. Over most of our parameter space (ie. \nel\ $\lesssim 10^4$
\cm3, cf. Fig.~\ref{fig:rmsden}), \neionic\ and \arionic\ can be 
considered in the low density
limit. The correction factor for \sionic\ is not larger than  1.35.
\nionic\ can be overestimated by up to a factor of 3.  However, the
strong dependence of \nionic\ on \nel\ can be circumvented by the use
of the LWS \OIII\ densities, which  characterize the ionic gas in the
same region where the nitrogen  lines are produced.

       \begin{figure}[!ht]
         \vspace{-0.55cm}
         \begin{center}
           \centerline{\psfig{file=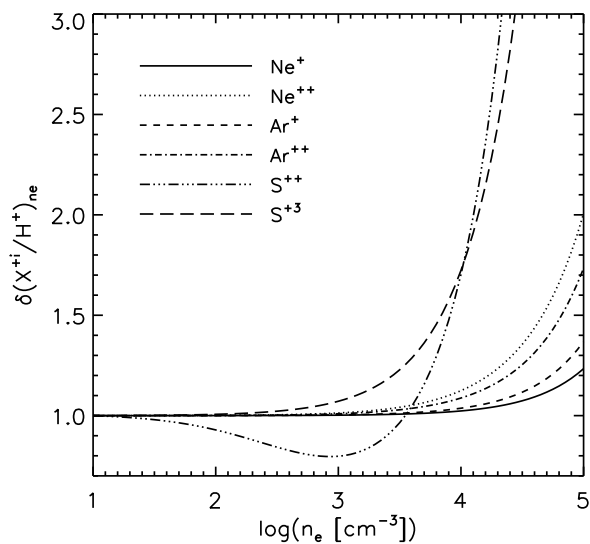,width=8.5cm}}
         \end{center}
         \vspace{-1.5cm}
         \caption{The \nel\ correction factor 
           $\delta(X^{+i}/{\rm H^+})_{n_{\rm e}}$ needed to apply to the 
           ionic abundance $X^{+i}/{\rm H^+}$ when the line
           emission coefficients are evaluated in the low density
           limit is plotted as a function of \nel.}
         \label{fig:abun_density}
       \end{figure}

To derive the ionic abundances with respect to hydrogen, one needs the
H$^+$ emission associated with the nebula.  These can be derived from
a near-infrared \HI\  recombination line, for instance Br$\alpha$,
which is detected in most of the sources, using
equation~(\ref{eq:ionicab}) as follows:

\begin{equation}
{ {X^{+i} \over {\rm H^{+}}} } 
={ { F_{X^{+i}} / F_{\rm Br_\alpha} } \over 
{ \epsilon_{X^{+i}} / \epsilon_{\rm Br_\alpha} } }~.
\label{eq:abundance_hi}
\end{equation}

The \HI\ emission coefficients show a dependence on electron
temperature that goes as \Tel$^{-\alpha}$ \citep[$\alpha=1.23$ for
Br$\alpha$,][]{hummer87}.  The electron temperature in \HII\
regions has been shown to increase with the galactocentric distance
\citep{shaver83,afflerbach96,afflerbach97,deharveng00}
from approximately 5000~K in the Galactic
Center to  10000~K at 15 kpc. Hence, if a single electron temperature
of 7500~K is assumed, the Galactic \Tel\ gradient will introduce a
systematic bias in the abundance determinations.  We can introduce a
\Tel\ correction factor  $\delta(X^{+i}/{\rm H^+})_{T_{\rm e}}$,
defined as $(X^{+i}/{\rm H^+})_{T_{\rm e}}/(X^{+i}/{\rm H^+})_{\rm
7500\,K}$ $\approx$ (\Tel/7500)$^{-1.23}$.  This correction factor
goes from $\sim 1.6$ at 5000~K to $\sim 0.7$ at 10000~K.

The H$^+$ emission can also be derived from optically thin, radio
continuum observations. Combining equations ~\ref{eq:flux_bralpha} and
~\ref{eq:abundance_hi}:

\begin{equation}
{ {X^{+i} \over {\rm H^{+}}} }= 1.03\times 10^{-14} {T_{\rm e}}^{-0.35}
\left ( {F_{X^{+i}} \over \epsilon_{X^{+i}}} \right ) 
\left ( {  {\nu^{-0.1}} \over S_\nu } \right ) 
\left ( { n_{\rm e} \over n_{\rm p} }\right ),
\label{eq:abundance_radio}
\end{equation}

\noindent
where $F_{X^{+i}}$ is in units of erg\,s$^{-1}$\,cm$^{-2}$,
$\epsilon_{X^{+i}}$ in erg\,s$^{-1}$\,cm$^3$,  \Tel\ in K, $\nu$ in
GHz and $S_\nu$ in Jy. The ratio of electrons to protons, $n_{\rm
e}/n_{\rm p}$, differs from unity because of the contribution of
singly ionized He to the total number of electrons. This can be
included by using   n$_{\rm e}$/n$_{\rm p} \approx$1+N$_{\rm
He^+}$/N$_{\rm H^+} \approx 1.10$. As can be seen from
equation~\ref{eq:abundance_radio},  this method shows a small
dependence on \Tel\ and the assumption of a single \Tel=7500~K gives
correction factors of the order of 15\% at most.

The dependence on \nel\ of the fine-structure emission coefficients
can also produce systematic uncertainties in the derived ionic
abundances with respect to H$^+$.  Fig.~\ref{fig:abun_density}
illustrates the case of the neon, argon and sulphur ionic abundances
(calculated using the lines \NeII\ 12.8, \NeIII\ 15.5, \ArII\ 7.0,
\ArIII\ 9.0, \SIII\ 18.7  and  \SIV\ 10.5~\mum)
and plots  the     \nel\ correction factor, $\delta(X^{+i}/{\rm
H^+})_{n_{\rm e}} \approx  [\epsilon_{X^{+i}}(n_{\rm e})/
\epsilon_{X^{+i}}(n_{\rm e} \rightarrow 0)]^{-1}$,  needed to apply
when the low density limit  is assumed. This correction becomes larger
than 1 when \nel\ $> n_{\rm crit}$. For the neon and argon abundances,
this correction only becomes important for \nel$> 10^4$ \cm3. More
critical is the case of the sulphur abundances because of the lower
critical densities of the \SIII\ 18.7 \mum\ ($10^4$ \cm3) and \SIV\
10.5 \mum\ ($3.7 \times 10^4$ \cm3) lines. At densities around $10^4$
\cm3, the sulphur abundance will be underestimated by a factor of 2.
Even more dramatic, although not shown in
Fig.~\ref{fig:abun_density}, is the case of N$^+$/H$^+$, N$^{++}$/H$^+$
and O$^{++}$/H$^+$. As in the case of \nionic, their strong \nel\
dependence can be  circumvented by the use of the LWS \OIII\ densities.

The total abundance of a given element is calculated by adding the
contributions of the different ionic species of that element. However,
ionization correction factors (ICFs) must be applied when some of the
ionic species are not observed.  The nebular abundances are then
determined by:
        
\begin{equation}
  {{X \over {\rm H}}} \simeq \sum_{i >0 } {{X^{+i} \over {\rm H^+}}} =
  {\rm ICF_{\rm unobserved~ions}} \times \sum_{\rm observed\ \it i} 
  {{X^{+i} \over {\rm H^+}}}~.
\end{equation}

In the case of the ISO spectral data, all the elements except oxygen
are observed in two ionization stages (see Fig.~1) and the ionization
corrections will therefore be close to unity for Ar, Ne, S and N. In
the case of oxygen, the abundance can only be derived by applying the
ICF for the unseen O$^+$

\subsection{Photoionization models}
\label{Models}

In order to evaluate in detail the dependence of the atomic
fine-structure line flux ratios of the \HII\ regions on the physical
conditions (stellar and nebular properties), a set of photoionization
models based on the code NEBU \citep{morisset96} was
performed.   For the stellar spectral energy distribution (SED), the
CoStar atmosphere models of \cite{schaerer97} were
used. These non-LTE stellar models include the effects of stellar wind
and line blanketing, important for massive stars.  A total of 26
CoStar models are available with effective temperatures ranging from
22 to 55~kK and luminosities from $1.5 \times 10^{38}$ to $8 \times
10^{39} \, \rm erg~s^{-1}$.  For each stellar atmosphere model, a
series of photoionization models was build for nebular densities of $3
\times 10^2$, $10^3$,  $3 \times 10^3$, $10^4$, $3 \times 10^4$ and
$10^5$ \cm3, which cover the density range probed by the atomic
fine-structure lines detected in the ISO spectra.  The nebulae are
assumed to be spherically symmetric. The nebular abundances are solar
\citep{grevesse98}, the distance between the star and the
nebula is taken to be $\rm 3 \times 10^{17} \, \rm cm$ and the
filling factor of the gas is supposed to be unity.   In total, 156
nebular models were calculated spanning a wide range in stellar
properties (one order of magnitude in luminosity) and nebular
densities (two orders of magnitude), corresponding to a range in the
ionization parameter of more than 3 orders of magnitude.  This large
range in ionization parameter combined with the effective stellar
temperatures is likely to describe the physical conditions of the
sample of \HII\ regions.  Since the main results deal with the
predictions of line flux ratios, the distances to the objects are not
relevant.  A detailed comparison of the predictions of such a model
with the ISO spectrum of the compact \HII\ region IRAS~18434$-$0242
(G29.96$-$0.02) is given in \cite{morisset01} - hereafter
Paper~III.

\section{Ionization state}
\label{Ionization}

Any two successive stages of ionization {\it i} and {\it i+1} of a
given element can be used as a measure of the state of ionization of
the nebula.  The line ratios \ArIII/\ArII\ 9.0/7.0, \NIII/\NII\
57.3/121.7,  \SIV/\SIII\ 10.5/18.7 and \NeIII/\NeII\ 15.5/12.8 \mum\
can be used to determine (cf. equation ~\ref{eq:ionicab})  the
relative ionic abundance ratios   \arionic, \nionic, \sionic\ and
\neionic, respectively.  A single \Tel=7500\,K and the low density
limit (except for \nionic, which was evaluated using the \OIII\
densities calculated in Sect.~\ref{Density}) are assumed in the
derivation.  The results  are presented in
Table~\ref{table:ionization}, assuming no correction for beam
differences or for differential extinction.  As mentioned in
Sect.~\ref{Methodology}, the low density limit is valid for \neionic\ and
\arionic\ over the range of the derived \nel(\OIII) and \rmsnel\
(\nel\ $\le 10^4$ \cm3), while \sionic\ may be affected by up to a 35\%
uncertainty.     The errors listed in Table~\ref{table:ionization}
stem from  the direct propagation of the line flux uncertainties (the
\nionic\ analysis includes also the uncertainties in the \OIII\
density). \neratio\ is not affected by calibration errors because the
lines involved in the ratio are measured in the same spectral band
(see Paper~I), resulting in smaller errors.
  
\begin{table}[!ht]
\begin{center}
\caption{Ionic abundances X$^{+i+1}$/X$^{+i}$ for argon, nitrogen,
  sulphur and neon derived in the low density limit (except for 
  \nionic, which was evaluated using the \OIII\ densities 
  calculated in Sect.~\ref{Density}).}
\label{table:ionization}
\leavevmode
\scriptsize
\begin{tabular}{lr@{\,}c@{\,}lr@{\,}c@{\,}lr@{\,}c@{\,}lr@{\,}c@{\,}lr@{\,}c@{\,}lr@{\,}c@{\,}lr@{\,}c@{\,}lr@{\,}c@{\,}l} \hline \\[1pt]
\multicolumn{1}{c}{Source}
& \multicolumn{3}{c}{Ar$^{++}$/Ar$^{+}$}
& \multicolumn{3}{c}{N$^{++}$/N$^{+}$}
& \multicolumn{3}{c}{S$^{+3}$/S$^{++}$}
& \multicolumn{3}{c}{Ne$^{++}$/Ne$^{+}$} \\[3pt] \hline \\[1pt]
IR\,01045 &   & --&  &   & --&  &   &  $<$ &  0.07 &   0.28 &  $\pm$ &   0.07 \\
IR\,02219 &   5.9 &  $\pm$ &   0.6 &   &  $>$ &  0.8 &   0.13 & $\pm$ &   0.02 &   1.07 &  $\pm$ &  0.06\\
IR\,10589 &   0.90 &  $\pm$ &   0.10 &   0.4 &  $\pm$ &   0.1 &   0.018 & $\pm$ &   0.004 &   0.13 &  $\pm$ &   0.01\\
IR\,11143 &   &  $>$ & 14 &   3 &  $\pm$ &   1 &   0.21 &  $\pm$ &   0.04 &   2.0 &  $\pm$ &   0.2\\
IR\,12063 &   4.6 &  $\pm$ &   0.6 &   &  $>$ &  1 &   0.15 &  $\pm$ &   0.03 &   0.88 &  $\pm$ &   0.06\\
IR\,12073 &   15 &   $\pm$ &   5 &   &  $>$ &  3 &   0.34 &  $\pm$ &   0.06 &   2.7 &  $\pm$ &   0.3\\
IR\,12331 &   2.1 &  $\pm$ &   0.3 &   0.8 &  $\pm$ &   0.2 &   0.027 &  $\pm$ &   0.005 &   0.32 &  $\pm$ &   0.02\\
IR\,15384 &   0.7 &  $\pm$ &   0.1 &   0.5 &  $\pm$ &   0.2 &   0.020 &  $\pm$ &   0.004 &   0.14 &  $\pm$ &   0.02\\
IR\,15502 &   0.24 & $\pm$ &   0.04&   0.5 &  $\pm$ &   0.2 &   0.009 &  $\pm$ &   0.002 &   0.11 &  $\pm$ &   0.02\\
IR\,16128 &   & --&  &   0.7 &  $\pm$ &   0.2 &   & --&  &   & --& \\
IR\,17160 &   & --&  &   0.31 &  $\pm$ &   0.09 &   & --&  &   & --& \\
IR\,17221 &   0.28 &  $\pm$ &   0.04 &   0.16 &  $\pm$ &   0.05 &   0.004 &  $\pm$ &   0.001 &   0.018 &  $\pm$ &   0.004\\
IR\,17279 &   & --&  &   0.24 &  $\pm$ &   0.06 &   & --&  &   & --& \\
Sgr\,C &   &  $<$ &  0.04 &   0.10 &  $\pm$ &   0.03 &   &  $<$ &  0.01 &  &  $<$ &  0.02\\
IR\,17455 &   0.66 &  $\pm$ &   0.07 &   0.6 &  $\pm$ &   0.2 &   0.016 &  $\pm$ &   0.003 &   0.20 &  $\pm$ &   0.01\\
IR\,17591 &   & --&  &   0.4 &  $\pm$ &   0.1 &   & --&  &   & --& \\
IR\,18032 &   0.14 &  $\pm$ &   0.02 &   0.19 &  $\pm$ &   0.08 &   &  $<$ &  0.003 &   &  $<$ &  0.01\\
IR\,18116 &   0.12 &  $\pm$ &   0.01 &   0.23 &  $\pm$ &   0.07 &   &  $<$ &  0.002 &   0.033 &  $\pm$ &   0.008\\
IR\,18317 &   0.22 &  $\pm$ &   0.02 &   0.10 &  $\pm$ &   0.04 &   0.003 &  $\pm$ &   0.001 &   0.029 &  $\pm$ &   0.002\\
IR\,18434 &   0.56 &  $\pm$ &   0.08 &   0.4 &  $\pm$ &   0.1 &   0.018 &  $\pm$ &   0.004 &   0.12 &  $\pm$ &   0.01\\
IR\,18469 &   & --&  &   0.4 &  $\pm$ &   0.1 &   & --&  &   & --& \\
IR\,18479 &   &  $<$ &  0.1 &   &  $>$ &  0.2 &   &  $<$ &  0.01 &   0.21 &  $\pm$ &   0.03\\
IR\,18502 &   0.38 &  $\pm$ &   0.06 &   &  $>$ &  0.4 &   0.011 &  $\pm$ &   0.003 &   0.100 &  $\pm$ &   0.008\\
IR\,19207 &   & --&  &   1.3 &  $\pm$ &   0.4 &   & --&  &   & --& \\
IR\,19442 &   &  $<$ &  0.05 &   & --&  &   &  $<$ &  0.03 &   &  $<$ &  0.01\\
IR\,19598 &   1.4 &  $\pm$ &   0.2 &   &  $>$ &  0.4 &   0.09 &  $\pm$ &   0.02 &   1.04 &  $\pm$ &   0.04\\
DR\,21 &   0.12 &  $\pm$ &   0.03 &   & --&  &   &  $<$ &  0.009 &   0.12 &  $\pm$ &   0.05\\
IR\,21190 &   2.9 &  $\pm$ &   0.4 &   & --&  &   0.06 &  $\pm$ &   0.01 &   0.43 &  $\pm$ &   0.02\\
IR\,21270 &   & --&  &   &  $>$ &  0.7 &   &  --  &       &   &   -- &\\
IR\,22308 &   0.23 &  $\pm$ &   0.08 &   &  $>$ &  0.2 &   &  $<$ &  0.008 &   &  $<$ &  0.02\\
IR\,23030 &   2.0 &  $\pm$ &   0.3 &   0.5 &  $\pm$ &   0.2 &   0.016 &  $\pm$ &   0.003 &   0.063 &  $\pm$ &   0.003\\
IR\,23133 &   0.25 &  $\pm$ &   0.03 &   & --&  &   0.004 &  $\pm$ &   0.001 &   &  $<$ &  0.01\\
\\ \hline
\end{tabular}
\end{center}
\vspace{-0.5cm}
\end{table}


       \begin{SCfigure*}[1.0][[!ht]
         \includegraphics[width=12.5cm]{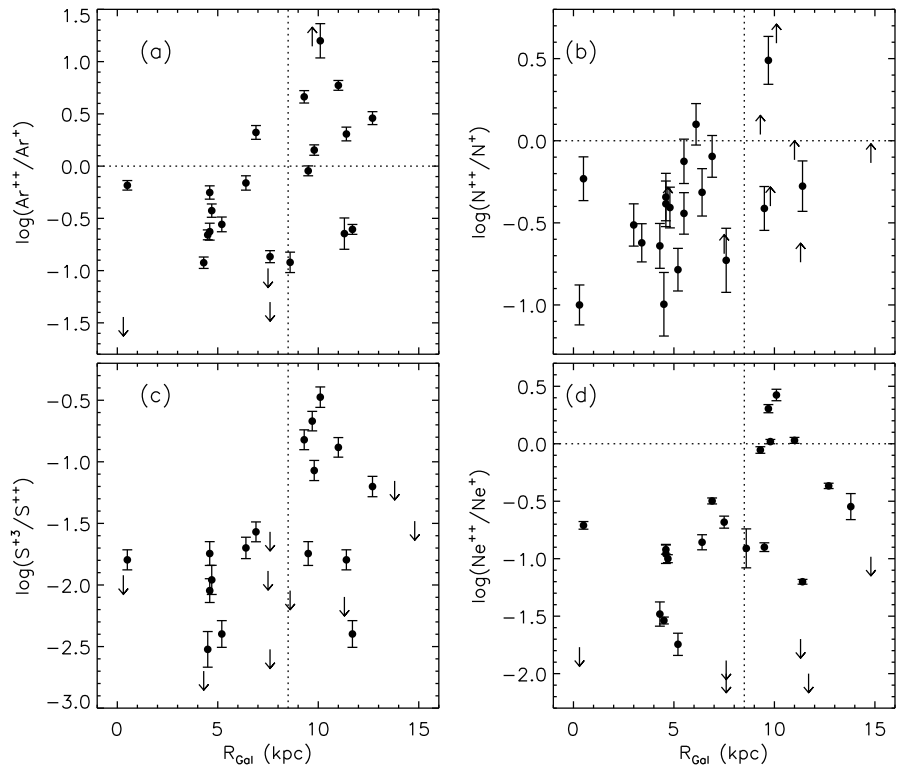}
         \hspace{-0.5cm}
         \caption{The spectral ``hardness'' indicators 
           \arionic, \nionic, \sionic\ and
           \neionic\ plotted against \rgal. 
           The dotted lines are intended only to guide the eye: the
           vertical line indicates the solar position at 8.5 kpc and
           the horizontal one indicates when $X^{+i+1}/X^{+i}=1$.
           The upper and lower limits are indicated by
           down and upwards arrows, respectively.
           \vspace{0.7cm}
           }
           \label{fig:ionizrgal}
       \end{SCfigure*}

Successive stages of ionization $X^{+i}$ and $X^{+i+1}$ of a given
element indicate the state of ionization of the nebula, which depends
only, in first approximation, on the ionization parameter, $U$,  and
the hardness of the ionizing radiation \citep{vilchez88}:

\begin{equation}
{ X^{+i+1} \over X^{+i}} \propto U {{\int_{\nu(X^{+i})}^{\infty}{
{J_\nu} \over {h\nu} } d\nu} \over {\int_{\rm 13.6\, eV}^{\infty}{
{J_\nu} \over {h\nu}} d\nu }}~,
\label{eq:ionization}
\end{equation}

\noindent
where $J_\nu$ is the mean intensity of the radiation.  Therefore, a
ratio $X^{+i+1}/X^{+i}$  is, to first order and for a given $U$,
proportional to the relative number of photons able to ionize
$X^{+i}$, ie. above $\nu(X^{+i})$, as compared to  that of Lyman
continuum photons (above 13.6 eV).  Hence, \arionic, \nionic, \sionic\
and \neionic\ probe the hardness of the stellar radiation  between
13.6 eV and 28, 30, 35 and 41~eV, respectively  (see
Fig.~\ref{fig:atomic}).

The behaviour of these ``hardness'' indicators with galactocentric
distance (cf. Fig.~\ref{fig:ionizrgal})  shows that the highest ionized
\HII\ regions are located at the highest galactocentric
distances. Assuming no systematic variations of $U$  with \rgal,
Fig.~\ref{fig:ionizrgal} indicates, consequently,  a hardening of the
radiation field towards large galactocentric distances.

\begin{SCfigure*}[1.0][!ht]
          \includegraphics[width=11.5cm]{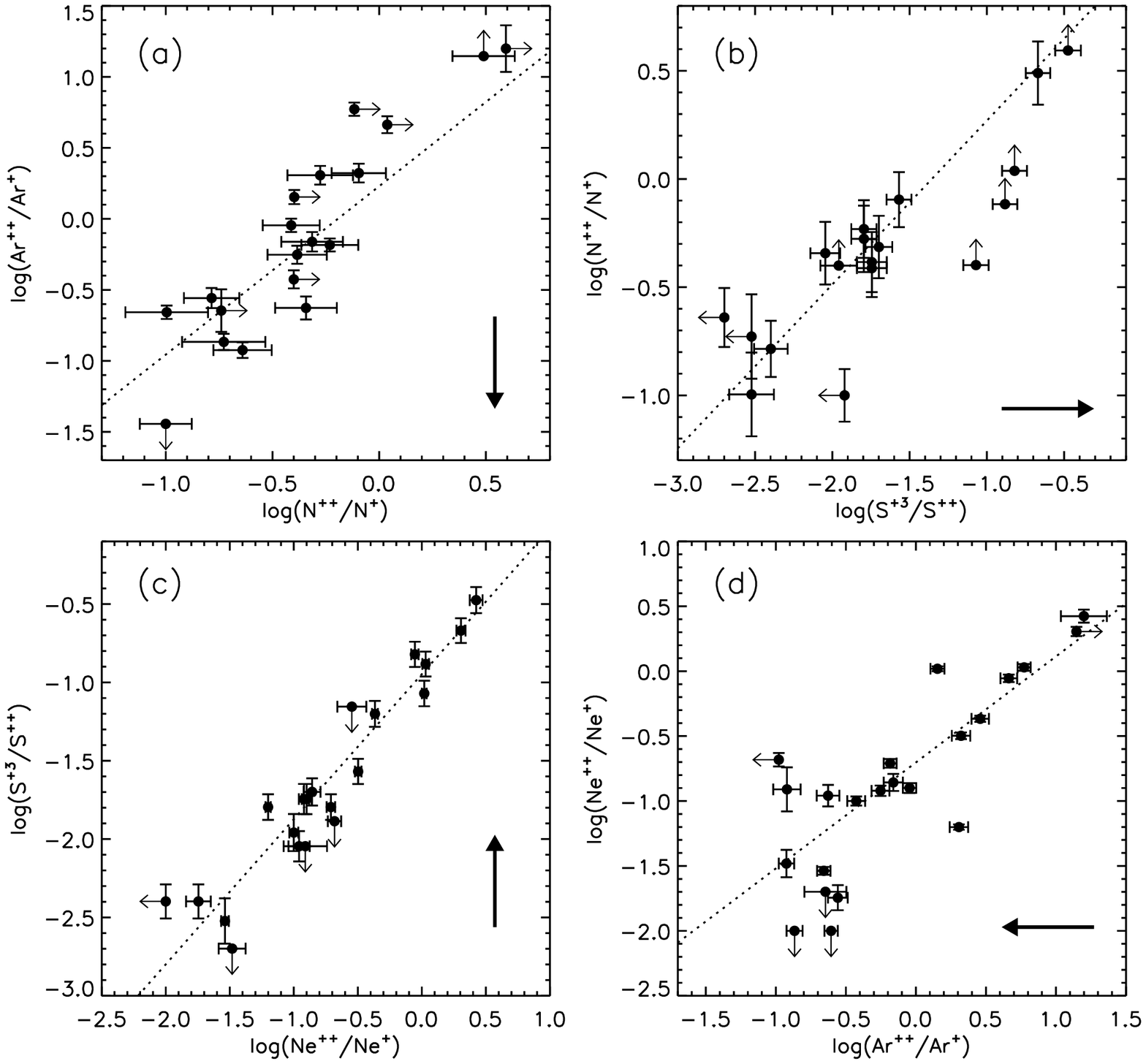}
          \caption{Correlations between 
            the ionic abundance ratios
            tracing the degree of ionization of the compact \HII\ regions:
            (a) \arionic\ vs. \nionic, (b) \nionic\ vs. \sionic\, (c)
            \sionic\ vs. \neionic\ and (d) \neionic\ vs.
            \arionic. Upper/lower limits are indicated by short
            arrows. The long arrows
            indicate the direction of extinction correction. If using
            the ``standard'' extinction quoted in
            Table~\ref{table:standard}, the extinction corrections to
            apply are: (a) 0.12$\times$\Ak, (b) 0.08$\times$\Ak, 
            (c) 0.08$\times$\Ak\ and (d) 0.12$\times$\Ak. 
            \vspace{0.4cm}
            } 
          \label{fig:ionizcross}
        \end{SCfigure*}

A ratio $(X^{i+1}/X^{+i})/(Y^{i+1}/Y^{+i})$ involving two different
elements $X$ and $Y$ measures to first order, according to
equation~\ref{eq:ionization}, the relative number of ionizing photons
in the respective ionizing continua \cite[e.g.][]{vilchez88},  
ie.\ above $\nu(X^{+i})$ and $\nu(Y^{+i})$, respectively.
Fig.~\ref{fig:ionizcross} shows  the correlations between the 4
``hardness'' indicators, ie. log(\arionic) vs. log(\nionic),
log(\nionic) vs. log(\sionic), log(\sionic) vs. log(\neionic) and
log(\neionic) vs. log(\arionic).  The arrows indicate the extinction
correction to be applied if the ``standard'' extinction law  (see
Sect.~\ref{Extinction:HI}) is assumed. The extinction correction for the
neon and nitrogen line ratios is negligible.  The dotted line in every
panel represents a straight line fit to the data and  is a fairly good
representation of the data behaviour.  The slopes obtained from a
least squares fit are very close to one: 1.2$\pm$0.3, 0.76$\pm$0.07,
0.93$\pm$0.06 and 0.8$\pm$0.1, respectively for panels {\it a}, {\it
b}, {\it c} and {\it d}.  Because these ionic pairs probe the ionizing
energies up to 41 eV,  Fig.~\ref{fig:ionizcross} shows that the
hardening of the  radiation field affects equally the full range of
the ionizing spectra.

\begin{figure}[!ht] 
          \vspace{-0.55cm}
          \psfig{figure={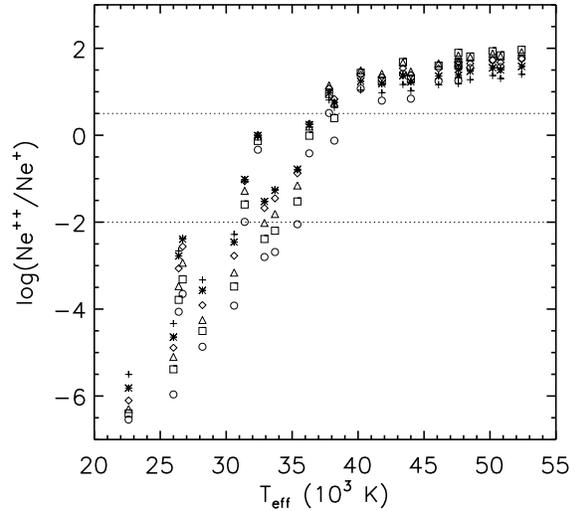},width=8.0cm}
          \vspace{-0.5cm}
        \caption{Model predictions for \neionic\  as
        a function of stellar effective temperature (\Teff) and
        nebular density. The densities are indicated by various
        symbols: the plus sign, star, diamond, triangle, square and
        circle correspond to
        densities of $3 \times 10^2$, $10^3$, $3 \times 10^3$,
        $10^4$, $3 \times 10^4$ and $10^5$ \cm3,
        respectively. The observed range for \neionic\ 
        ($\sim$ 0.01 to 3) in the ISO sample of \HII\ 
        regions is delineated by the dotted horizontal lines.
        \vspace{-0.5cm}} 
           \label{fig:exne_teff} 
        \end{figure}

Photoionization models, however,  show an additional dependence of
$(X^{+i+1}/X^{+i})/(Y^{+i+1}/Y^{+i})$ on $U$ \citep[][Paper~III]{stasinska97}  
rendering the interpretation more complex and
probably producing part of the scatter in Figs.~\ref{fig:ionizrgal}
and ~\ref{fig:ionizcross}.  The set of photoionization models
described in Sect.~\ref{Models} allows us to study the influence of the
ionization parameter $U$ and the stellar effective temperature \Teff\
(ie. the shape of the ionizing spectrum) on the ionization state of
the gas, traced by any of the ratios $X^{+i+1}/X^{+i}$, for instance
\neionic.  The variation of \neionic\ is shown as a function of
\Teff\ in Fig.~\ref{fig:exne_teff}.   For a given stellar atmosphere
model or \Teff, this ratio decreases with increasing electron
density. The observed \neionic\ ranges from $\sim$ 0.01 to 3, a range
predicted by the models for values of \Teff\ between 31 and 38~kK
(Fig.~\ref{fig:exne_teff}).  Similar variations are predicted for the
other ionization ratios (\nionic, \arionic\ and \sionic).  Changing
$U$ at a given \Teff\ (by changing the stellar luminosity  and/or the
gas density) can change \neionic\ by a factor of 10 or more.  An
estimate of the \Teff\ of the ionizing star is therefore only possible
if robust constraints on $U$ are available.   A detailed study of the
relations between the ionization diagnostics and the stellar
atmosphere models is postponed to a future paper (Morisset et al. in
prep.).

However, as tracers of the ionization state, the ionic abundance
ratios $X^{+i+1}/X^{+i}$ are sensitive indicators of the ionization
correction factors.  Fig.~\ref{fig:icfs} shows the ICFs for N, O, Ne,
S and Ar derived from the grid of photoionization models as a function
of \neionic.  The scatter in the calculated ionization correction
factors is low, indicating that the use of the ionic abundance ratios
$X^{+i+1}/X^{+i}$ to constrain the ICFs is a robust method. In other
words, even if $U$ and the SED of the ionizing star are poorly known,
the knowledge of the ionization state of the gas through the
ionization ratios is enough to determine reliable ICFs. This
conclusion does not strongly depend on the stellar atmosphere models
as shown by comparing the results of a photoionization code using
other atmosphere models (e.g. CMFGEN - Bourret, private
communication).

The observed values of the ionization ratios can therefore be used to
correct the elemental abundances for the unseen ions. As shown in
Fig.~\ref{fig:icfs}, the effects are rather small, except for oxygen.
The effect on the Ne abundance is negligible.  In the case of sulphur,
the abundance is always underestimated by typically $\sim$~15\%.   For
high ionization conditions, the abundance of N and Ar can be
underestimated by up to 10 and 30\%, respectively, because of the
missing N$^{+3}$ and Ar$^{+3}$. For O, only O$^{++}$ is available and
thus, we are missing the low ionization  state O$^+$. The ICF for O is
very large, between 1.5 and 100.

\section{Abundances}
\label{Abundances}

\subsection{\noionic\ and N/O}
\label{N/O}

The \noionic\ abundance ratio is straightforward to derive from
equation~\ref{eq:ionicab} using the \NIII\ 57~\mum\ and \OIII\
52$+$88~\mum\ line fluxes together with the densities derived from
the \OIII\ 88/52 line ratio (Table~\ref{table:o3den}), which are  well
adapted because both the N and O lines are emitted by low density
gas. All three lines are observed with LWS and consequently, beam
differences and differential extinction are negligible.  Finally, the
use of both \OIII\ lines minimizes considerably the dependence of the
ionic abundance on \nel.  The resulting \noionic\ ionic abundance
ratios are listed in Table~\ref{table:noionic} (the quoted
uncertainties come from the error propagation of the line fluxes).
The ionic abundance ratio $\rm N^{++}/O^{++}$ is found to decrease
with galactocentric distance (Fig.~\ref{fig:noionicrgal}).

       \begin{figure}[!ht] 
          \vspace{-0.8cm}
        \psfig{figure={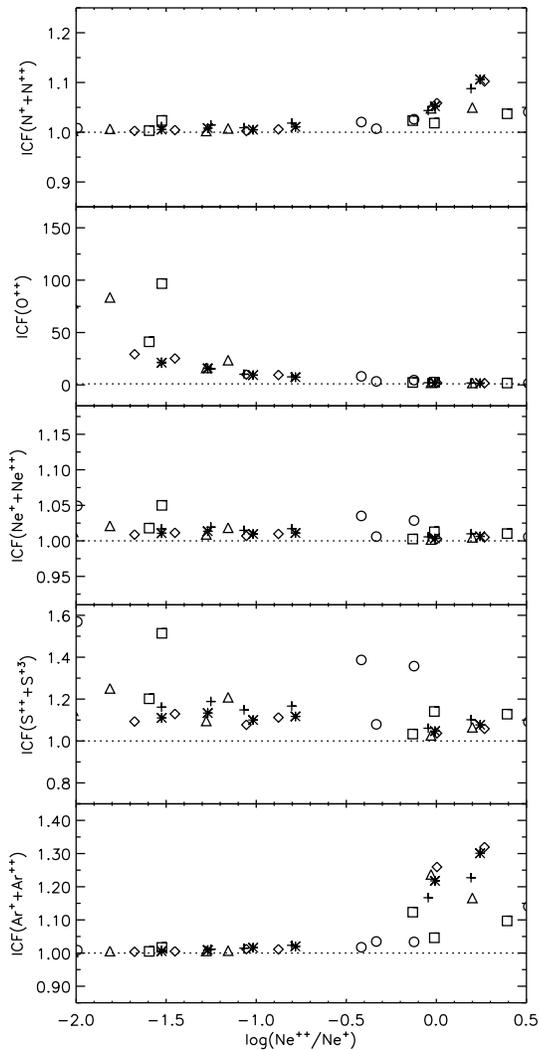},width=8.5cm}
        \vspace{-0.5cm}
        \caption{Model predictions for the ionization correction
        factors (ICFs) for N, O, Ne, S and Ar. The ICFs are plotted
        against \neionic, which traces the ionization state. The
        dotted line indicates an ICF equal to one. The densities are
        indicated by various symbols: $3 \times 10^2$ \cm3\ (plus signs),
        $10^3$ \cm3\ (stars), $3 \times 10^3$ \cm3\ (diamonds),
        $10^4$ \cm3\ (triangles), $3 \times 10^4$ \cm3\ (squares) and
        $10^5$ \cm3\ (circles).}
        \label{fig:icfs} 
        \end{figure}

\begin{table}[!ht]
\caption{The ionic abundance ratio N$^{++}$/O$^{++}$ as derived from the LWS fine-structure lines \NIII\ 57~\mum\ and \OIII\ 52+88~\mum.}
\label{table:noionic}
\begin{center}
\leavevmode
\footnotesize
\begin{tabular}{lr@{\,}c@{\,}llr@{\,}c@{\,}l} \hline \\[1pt]
\multicolumn{1}{c}{Source}
& \multicolumn{3}{c}{N$^{++}$/O$^{++}$}
& \multicolumn{1}{c}{Source}
& \multicolumn{3}{c}{N$^{++}$/O$^{++}$} \\[2pt] \hline \\[1pt]
IR\,02219 &  0.16 & $\pm$ &   0.04&IR\,18032 &    0.6 & $\pm$ &    0.1\\
IR\,10589 &  0.24 & $\pm$ &   0.04&IR\,18116 &   0.50 & $\pm$ &   0.08\\
IR\,11143 &  0.14 & $\pm$ &   0.02&IR\,18317 &    0.8 & $\pm$ &    0.1\\
IR\,12063 &  0.18 & $\pm$ &   0.03&IR\,18434 &   0.49 & $\pm$ &   0.07\\
IR\,12073 &  0.13 & $\pm$ &   0.02&IR\,18469 &   0.43 & $\pm$ &   0.07\\
IR\,12331 &  0.30 & $\pm$ &   0.04&IR\,18479 &   0.15 & $\pm$ &   0.04\\
IR\,15384 &  0.41 & $\pm$ &   0.06&IR\,18502 &   0.39 & $\pm$ &   0.07\\
IR\,15502 &  0.47 & $\pm$ &   0.08&IR\,19207 &   0.24 & $\pm$ &   0.04\\
IR\,16128 &  0.38 & $\pm$ &   0.06&IR\,19598 &   0.10 & $\pm$ &   0.02\\
IR\,17160 &  0.50 & $\pm$ &   0.07& DR\,21 &   &$<$ &0.21   \\
IR\,17221 &   0.7 & $\pm$ &    0.1&IR\,21190 & &$<$ &0.33    \\
IR\,17279 &   0.7 & $\pm$ &    0.1&IR\,21270 &   0.17 & $\pm$ &   0.03\\
 Sgr\,C &   0.7 & $\pm$ &    0.1&IR\,21306 &   &$<$ &0.13    \\
IR\,17455 &  0.44 & $\pm$ &   0.07&IR\,22566 &    0.2 & $\pm$ &    0.1\\
IR\,17591 &  0.12 & $\pm$ &   0.02&IR\,23030 &   0.33 & $\pm$ &   0.05\\
IR\,18032 &   0.6 & $\pm$ &    0.1\\
\\ \hline
\end{tabular}
\end{center}
\end{table}


The significance of \noionic\ and its dependence on \rgal\ has been
discussed in detail in previous studies \citep[e.g.][and
references therein]{rubin88}. \noionic\ is a function of both the ionization
state of the nebula (the ionization potentials of N$^{+}$ and O$^{+}$
are 29.6 and 35.1~eV, respectively) as well as of abundance, and one
has to  disentangle their effects to derive the N/O elemental
abundance ratio.  Fig.~\ref{fig:ioniz_noionic} presents the
correlation between   \noionic\ and \neionic\ (note that the spectral
``hardness'' indicators are inter-correlated and thus, they all
correlate with \noionic).  As can be seen in
Figs. ~\ref{fig:noionicrgal} and  ~\ref{fig:ioniz_noionic}, \noionic\
is inversely proportional to  the degree of ionization (ie. low
ionized nebulae have high \noionic\ ionic  abundances) and this
dependence is connected to the location in the Galaxy (ie. nebulae
with high \noionic\ are preferentially located in the inner Galaxy).
For these inner nebulae, N$^{+}$ and  O$^{+}$ are probably the
dominant ionic species over  N$^{++}$ and  O$^{++}$ and therefore, the
ionization correction  needed to get N/O from \noionic\ could be
rather large (\noionic\ will be then overestimating N/O). In contrast,
\noionic\ must be close to N/O for  the nebulae in the outer Galaxy.

The predictions of the photoionization models are useful to quantify
the effects of ionization conditions on \noionic.   In particular, one
can estimate the ionization correction factor which has to be applied
by using one of  the ionization diagnostic $X^{+i+1}/X^{+i}$.  Based on
the model results, an empirical  method can be determined to derive
the elemental abundance ratio  N/O from the observed \NIII\ and \OIII\
line fluxes.  The N/O abundance ratio can be expressed as follows:

\begin{equation}
{\rm N \over O}= {[\ion{N}{iii}] \over {\Sigma[\ion{O}{iii}]}} \times
 {{\rm N^{++} / {\rm O^{++}}} \over 
{[\ion{N}{iii}] / {\Sigma[\ion{O}{iii}]}}} \times
{{\rm N / O} \over {\rm N^{++} / {\rm O^{++}}}}~,
\label{N/O}
\end{equation}

\noindent
where the first term on the right side is the ratio of the \NIII\ and
\OIII\ line fluxes, the second term is the ratio of the emissivities
of these lines (taking into account the density effect leading to
collisional de-excitation) and the third term is the ratio of the ICFs
which have to be applied to N and O. $\Sigma$[\ion{O}{iii}] is the sum
of the  two \OIII\ lines, ie. \OIII~52+88~\mum.

The second and third member of the equation~\ref{N/O} can be
empirically  determined by the use of appropriate observables and the
grid of models  leading to the following relation:

\begin{equation}
\begin{array}{r}
{\rm N \over O} = {[\ion{N}{iii}] \over {\Sigma[\ion{O}{iii}]}}\times 
\left ( 1.07 r[\ion{O}{iii}]^{-0.29} \right ) \\[5pt]  
\times \left (0.95+0.27 {\rm log}{[\ion{Ne}{iii}] \over 
{[\ion{Ne}{ii}]}} \right ),
\label{eq:method_ne}
\end{array}
\end{equation}

\noindent
where r\OIII\ is the ratio \OIII~88/52~\mum. The density correction
via r\OIII, which transforms   [\ion{N}{iii}]/$\Sigma$[\ion{O}{iii}]
into \noionic, yields \noionic\ ratios on average 10\% lower than the
\noionic\ derived from equation~\ref{eq:ionicab},  mainly due to
slight differences between  the atomic constants used in the code NEBU
and the ones used in this work.

       \begin{figure}[!ht]
         \vspace{-0.55cm}
         \begin{center}
           \centerline{\psfig{file=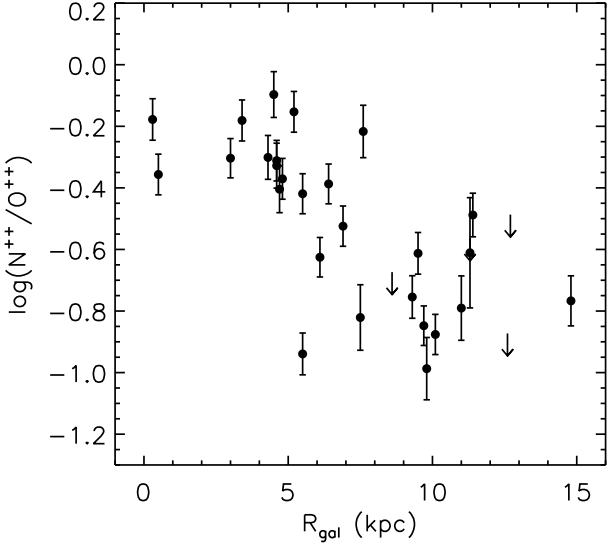,width=8.5cm}}
         \end{center}
         \vspace{-1.5cm}
         \caption{Plot of the ionic abundance ratio N$^{++}$/O$^{++}$ versus 
           the galactocentric distance \rgal. The upper limits are
           denoted by downwards arrows. \vspace{-0.3cm}} 
         \label{fig:noionicrgal}
       \end{figure}

The photoionization model predictions of these successive ratios are
shown in Fig.~\ref{fig:neon_corr} as a function of degree of
ionization  (as measured by the ionic abundance ratio \neionic): the
line flux ratio  [\ion{N}{iii}]/$\Sigma$[\ion{O}{iii}] (plus signs),
the ionic abundance  ratio \noionic\ (diamonds)  and the elemental
abundance ratio N/O after correcting for the density and the unseen
ions.  The dotted horizontal line indicates the input value of N/O of
the model, ie. the solar ratio of 0.123 \citep{grevesse98}.
As can be seen, when no ICF corrections are applied, the \noionic\
ratio overestimates the N/O abundance ratio for low ionization
conditions and slightly underestimates the N/O ratio for high
ionization conditions.  Finally, the results for the N/O abundance
ratio after correcting for the density and the ICF dependence (stars
in Fig.~\ref{fig:neon_corr}) agree very well with the input solar
value of 0.123.

Equation~\ref{eq:method_ne} provides a useful empirical law to derive
the N/O abundance ratio from the observed \OIII, \NIII\ and neon line
fluxes. Any deviation from the solar N/O abundance after applying this
law to the observed [\ion{N}{iii}]/$\Sigma$[\ion{O}{iii}] indicates
intrinsic variations in the N/O abundance ratio.  Similar empirical
laws can be derived for the \arratio\ and \sratio\  ionization
diagnostics. However, we prefer the method based on the neon lines,
whose ratio is hardly affected by differential extinction and
aperture differences.  Note that the \nratio\ has an extra dependence
on density and is,  therefore, less appropriate.

Fig.~\ref{fig:norgal} shows the N/O abundance ratio as  a function of
galactocentric distance after applying the corrections described above
by using the neon lines. Note that there are only 25 sources for which
the corrections could be applied. Whereas the slope of the variation
is not as steep as that for $\rm N^{++}/O^{++}$  (see
Fig.~\ref{fig:noionicrgal}), a gradient of N/O  with galactocentric
distance is clearly present in the sense of a decreasing ratio with
$\rm R_{Gal}$:

       \begin{figure}[!ht]
         \vspace{-0.55cm}
         \begin{center}
           \leavevmode
           \centerline{\psfig{file=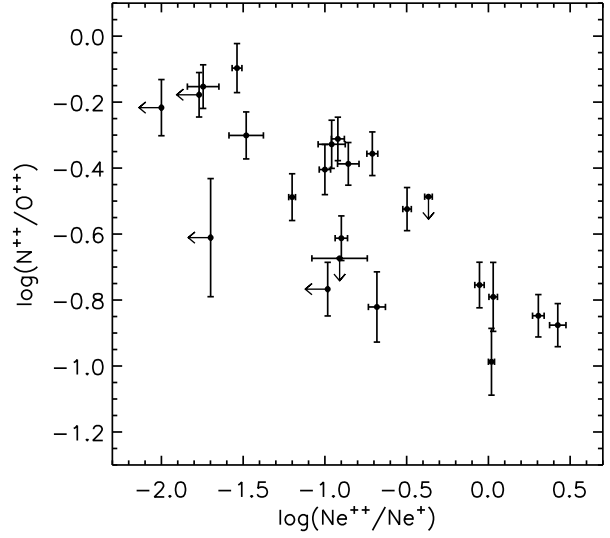,width=8.5cm}}
         \end{center}
         \vspace{-1.1cm}
         \caption{Plot of \noionic\ versus \neionic, a degree of
           ionization indicator.}
         \label{fig:ioniz_noionic}
       \end{figure}

        \begin{figure}[!ht] 
         \vspace{-0.55cm}
         \hspace{0.05cm}
         \psfig{figure={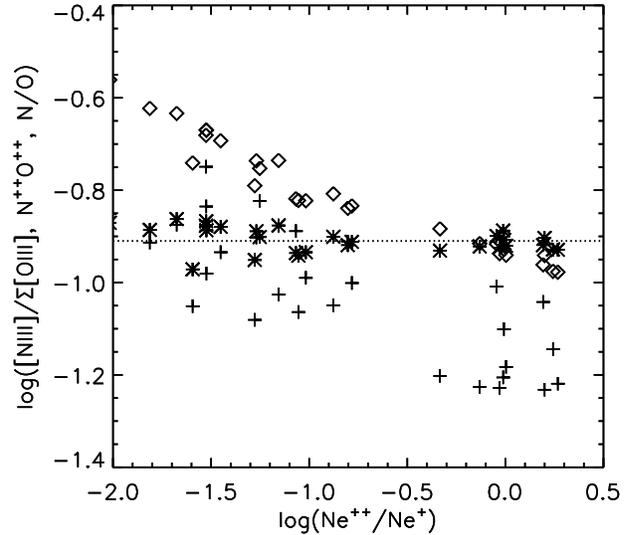},width=8.5cm}
         \vspace{-0.8cm}
        \caption{Model predictions for the variation of 
          [\ion{N}{iii}]/$\Sigma$[\ion{O}{iii}]        
          (plus signs) as a function of the ionization diagnostic
          \neionic\ $\propto$ \neratio.  The horizontal line indicates the 
          input
          value of N/O of the model, ie. the solar ratio of 0.123. 
          The ionic abundance ratios \noionic\
          (after correcting the line fluxes for the collisional de-excitation)
          are shown as diamonds. The N/O elemental abundance ratios (after 
          correcting for the density and the unseen ions) are shown as
          stars - see text.} 
        \label{fig:neon_corr}
        \end{figure}

        \begin{figure}[!ht] 
          \vspace{-0.55cm}
        \psfig{figure={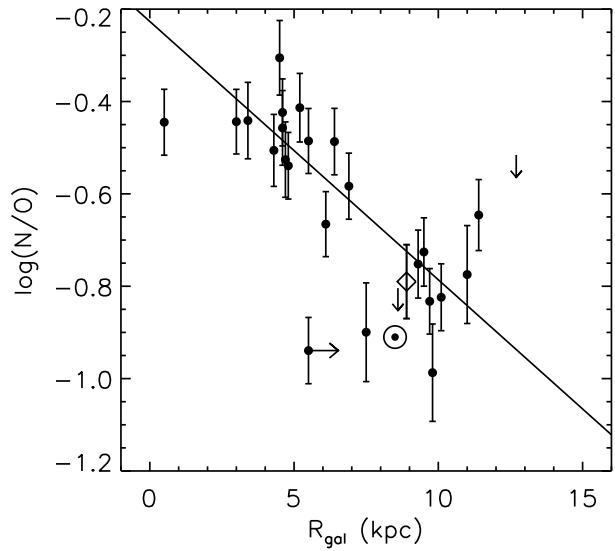},width=8.5cm}
         \vspace{-0.6cm}
        \caption{Plot of the elemental abundance ratio
        N/O versus galactocentric distance $\rm R_{Gal}$
        as derived from the $\rm N^{++}/O^{++}$ ratio - see
        text. The upper limits are denoted by downwards arrows.}
        \label{fig:norgal}
        \end{figure}

\begin{table}[!ht]
\caption{Ionic and total abundances derived from the LWS
  fine-structure lines. These abundances are very uncertain because of
the possible mismatch between the gas observed by LWS and the radio
emission used to derive H$^+$ (see Sect.~\ref{LWSabun}).}
\label{table:abunradio}
\begin{center}
\leavevmode
\scriptsize
\begin{tabular}{lr@{\,}c@{\,}lr@{\,}c@{\,}lr@{\,}c@{\,}lr@{\,}c@{\,}l}\hline \\[1pt]
\multicolumn{1}{c}{Source}
& \multicolumn{3}{c}{N$^{+}$/H$^{+}$}
& \multicolumn{3}{c}{N$^{++}$/H$^{+}$}
& \multicolumn{3}{c}{N/H}
& \multicolumn{3}{c}{O$^{++}$/H$^{+}$}\\
\multicolumn{1}{c}{ }
& \multicolumn{3}{c}{$\times 10^{-4}$}
& \multicolumn{3}{c}{$\times 10^{-4}$}
& \multicolumn{3}{c}{$\times 10^{-4}$}
& \multicolumn{3}{c}{$\times 10^{-4}$} \\[2pt] \hline \\[1pt]
IR\,02219 &  &  $<$ &  0.3 &   0.2 &  $\pm$ &   0.1 &  &  $<$ &  0.5 &   1.4
 &  $\pm$ &   0.6\\
IR\,17160 &   1.5 &  $\pm$ &   0.6 &   0.5 &  $\pm$ &   0.2 &   2.0 & 
 $\pm$ &   0.8 &   0.9 &  $\pm$ &   0.3\\
IR\,17279 &   9. &  $\pm$ &   4. &   2.0 &  $\pm$ &   0.6 &  11. & 
 $\pm$ &   4. &   3.0 &  $\pm$ &   0.9\\
IR\,17455 &   1.3 &  $\pm$ &   0.6 &   0.8 &  $\pm$ &   0.3 &   2.1 & 
 $\pm$ &   0.8 &   1.7 &  $\pm$ &   0.6\\
IR\,17591 &   0.3 &  $\pm$ &   0.1 &   0.12 &  $\pm$ &   0.04 &   0.5 & 
 $\pm$ &   0.2 &   1.0 &  $\pm$ &   0.3\\
IR\,18032 &   1.7 &  $\pm$ &   0.9 &   0.3 &  $\pm$ &   0.1 &   2. & 
 $\pm$ &   1. &   0.5 &  $\pm$ &   0.2\\
IR\,18116 &   0.5 &  $\pm$ &   0.2 &   0.11 &  $\pm$ &   0.04 &   0.6 & 
 $\pm$ &   0.2 &   0.21 &  $\pm$ &   0.07\\
IR\,18317 &   4. &  $\pm$ &   2. &   0.4 &  $\pm$ &   0.2 &   5. & 
 $\pm$ &   2. &   0.5 &  $\pm$ &   0.2\\
IR\,18434 &   1.3 &  $\pm$ &   0.6 &   0.5 &  $\pm$ &   0.2 &   1.9 & 
 $\pm$ &   0.8 &   1.1 &  $\pm$ &   0.4\\
IR\,18469 &   7. &  $\pm$ &   3. &   2.5 &  $\pm$ &   0.8 &   9. & 
 $\pm$ &   4. &   6. &  $\pm$ &   2.\\
IR\,18479 &  &  $<$ &  0.3 &   0.07 &  $\pm$ &   0.03 &  &  $<$ &  0.4 &   0.4
 &  $\pm$ &   0.2\\
IR\,18502 &  &  $<$ &  0.5 &   0.20 &  $\pm$ &   0.08 &  &  $<$ &  0.7 &   0.5
 &  $\pm$ &   0.2\\
IR\,19598 &  &  $<$ &  0.1 &   0.05 &  $\pm$ &   0.02 &  &  $<$ &  0.2 &   0.5
 &  $\pm$ &   0.2\\
 DR\,21 &  &  $<$ &  0.04 &  &  $<$ &  0.009 &   & --&  &   0.04 & 
 $\pm$ &   0.01\\
IR\,21190 &  &  $<$ &  0.7 &  &  $<$ &  0.2 &   & --&  &   0.6 & 
 $\pm$ &   0.3\\
IR\,23030 &   0.3 &  $\pm$ &   0.1 &   0.16 &  $\pm$ &   0.06 &   0.5 & 
 $\pm$ &   0.2 &   0.5 &  $\pm$ &   0.2\\
\\ \hline
\end{tabular}
\end{center}
\end{table}


\begin{equation}
{\rm log {N \over O} } = (-0.23 \pm 0.07) - (0.056 \pm 0.009) {\rm R_{Gal}}~.
\end{equation}

The correlation coefficient of the least square fit is r=$-0.80$.
This Galactic N/O gradient is consistent with previous studies of
\HII\ regions \citep{simpson95a} and  B stars \citep{smartt01}.

\subsection{N/H and $O^{++}/H^{+}$}
\label{LWSabun}

Because of the large difference between the SWS and LWS apertures and
the fact that many sources are more extended than the SWS aperture,
the hydrogen line flux corresponding to the gas where the  nitrogen
and oxygen lines are formed will be underestimated if a SWS \HI\
recombination line is used. Thus, the resulting abundances of N and O
will be overestimated.  Hence, the radio  continuum measurements are
thought to be more appropriate to evaluate H$^+$ and
equation~\ref{eq:abundance_radio} is used to calculate N$^{+}$/H$^+$,
N$^{++}$/H$^+$ and O$^{++}$/H$^+$ via the lines \NII\ 122, \NIII\
57 and \OIII\ 52+88 \mum. The radio flux densities from
Table~\ref{table:radio} are used and the sources which are more
extended than the LWS aperture  (Sgr\,C, IR\,19207, IR\,21270 and
IR\,22308) are excluded. \OIII\ densities
(cf. Table~\ref{table:o3den}) are used to evaluate the line
emissivities involved.

       \begin{figure}[!hb]
         \vspace{-0.8cm}
          \begin{center}
            \centerline{\psfig{file=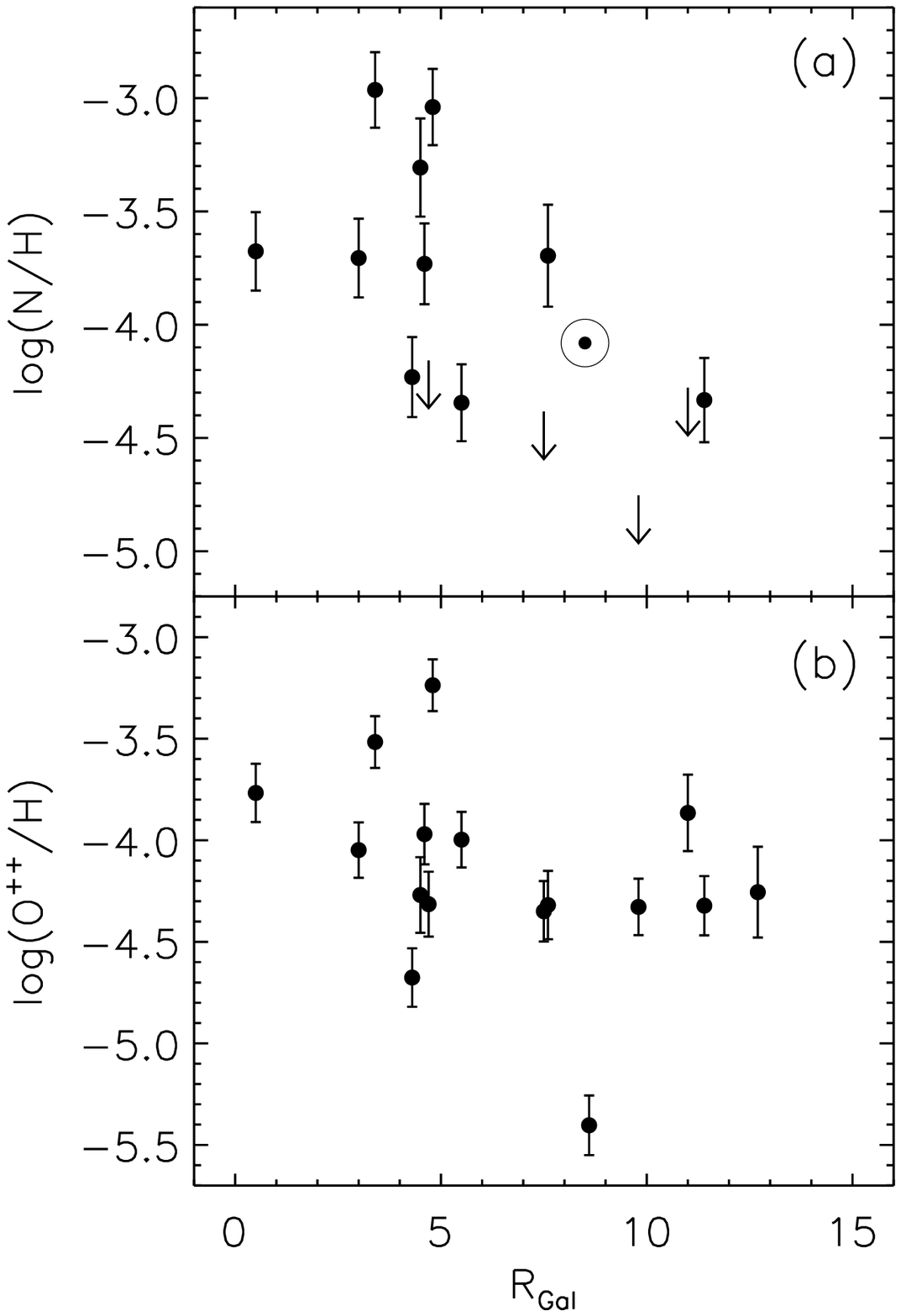,width=8.5cm}}
          \end{center}
          \vspace{-0.8cm}
          \caption{(a) N/H and (b) O$^{++}$/H$^+$ as a function of
            \rgal. The solar
            nitrogen abundance \citep{grevesse98} is indicated by the
            symbol \sun.}
          \label{fig:abunradio}
        \end{figure}

For nitrogen, both ions N$^{+}$ and N$^{++}$ are
measured. Fig.~\ref{fig:ionizrgal} shows that N$^{+}$ is the dominant
species over N$^{++}$ and therefore, N$^{+3}$ is not expected to
contribute significantly to the total abundance of nitrogen. Indeed,
the grid of photoionization models (cf. Fig.~\ref{fig:icfs}) shows that
the contribution of N$^{+3}$ for the highest ionized sources is only
at the 10\% level.  Hence, the total nitrogen abundance N/H is
calculated by adding the ionic abundances N$^{+}$/H$^{+}$ and
N$^{++}$/H$^{+}$. The calculated N$^{+}$/H$^{+}$, N$^{++}$/H$^{+}$
and N/H are given in
Table~\ref{table:abunradio}. Fig.~\ref{fig:abunradio}~(a) represents
N/H versus \rgal. An attempt to fit a straight line was made, but no
statistically significant fit was possible    because of the few
points and the large scatter. We note that  sources in the inner 5 kpc
have supersolar nitrogen abundances, in agreement with determinations
using B stars \citep[e.g.][]{smartt01}.

For oxygen, only O$^{++}$ is available and thus, we are missing
O$^{+}$. We have used the grid of photoionization models to estimate
the effect of the missing O$^+$ abundance. These models show that the
ionization correction factor is very large, between 1.5 and 100
depending on the degree of ionization (cf. Fig.~\ref{fig:icfs}).
Therefore, the derived O$^{++}$/H$^{+}$ listed in
Table~\ref{table:abunradio} can only be interpreted as a lower limit
for the total O/H abundance. Fig.~\ref{fig:abunradio}~(b) plots
O$^{++}$/H$^{+}$ versus \rgal. Clearly, the derived O$^{++}$/H$^{+}$
abundances are much lower than the solar log(O/H)$=-3.2$ \citep{grevesse98}.

A note of caution must be brought concerning the nitrogen and oxygen
abundances. In most cases, the ISO \OIII~88/52 \mum\ line ratio give
densities which are far below those required to produce the
ultracompact \HII\ emission observed at radio wavelengths (${\rm
n_{e,rms}}$) (see Fig.~\ref{fig:rmsden}). This suggests that the
ultracompact \HII\ regions are surrounded by a tenuous shell of
O$^{++}$ (and probably also N$^{++}$). If this is the case, both N and
O$^{++}$ abundances will not be the true nebular abundances as the
H$^+$ emission associated with the shell will not be the one given by
the radio observations.

\subsection{Ne/H, S/H and Ar/H} 
\label{Alpha}

Despite the inherent uncertainty associated with the \HI\
recombination line method concerning \Tel\
(cf. Sect.~\ref{Methodology}),  we will use the Br$\alpha$ line  fluxes
observed with ISO/SWS in analyzing all the SWS fine-structure lines of
neon, sulphur and argon because the beams are well matched.  As the
lines of \NeII\ 12.8 and \NeIII\ 15.5, \SIII\ 18.7 and \SIV\ 10.5, and
\ArII\ 7.0 and \ArIII\ 9.0, used to  determine the abundances of neon,
sulphur and argon, respectively, are  observed by a beam comparable to
Br$\alpha$, the use of equation
~\ref{eq:abundance_hi}  is straightforward.  The fine-structure lines
and \HI\ recombination line emissivities involved in the calculation
were evaluated in the low density limit and considering a single
\Tel=7500 K.  The lines in the SWS band 4, \SIII\ 33.4 and \NeIII\
36.0 \mum, which are  observed in a larger aperture and are affected
by large calibration uncertainties, are not included in the
determination (see  discussion in Sect.~\ref{Density}).

The absolute abundances for neon, sulphur and argon were calculated 
adding the contribution from the different ionization stages available:

\begin{equation}
{\rm Ne \over  H } = {\rm Ne^{+} \over H^+} + {\rm Ne^{++} \over H^+}~;  
\end{equation}  

\begin{equation}
{\rm S \over  H } ~= {\rm S^{++} \over H^+} + {\rm S^{+3} \over H^+}~;  
\end{equation}  

\begin{equation}
{\rm Ar \over  H } = {\rm Ar^+ \over H^+} + {\rm Ar^{++} \over H^+}~.  
\end{equation}  

The grid of photoionization models described in Sect.~\ref{Models}
allows us to verify the validity of the above expressions
(cf. Fig.~\ref{fig:icfs}). The ionization correction factor for Ne is
$\sim$~1 for all the sources. Ar$^{+3}$ may be present in the higher
ionized sources (photons with energies $>$ 40.7 eV  are necessary to
ionize Ar$^{++}$  to Ar$^{+3}$, roughly the same energies to produce
Ne$^{++}$). However, the contribution of this ion is estimated to be
$< 30 \%$. The S abundance is always underestimated by $\sim$ 15\%.
Table~\ref{table:alpha} gives the ionic and total elemental abundances
for argon, sulphur and neon. The uncertainties are derived propagating
the line flux errors.

\begin{figure}[!ht]
         \vspace{-0.8cm}
          \begin{center}
            \centerline{\psfig{file=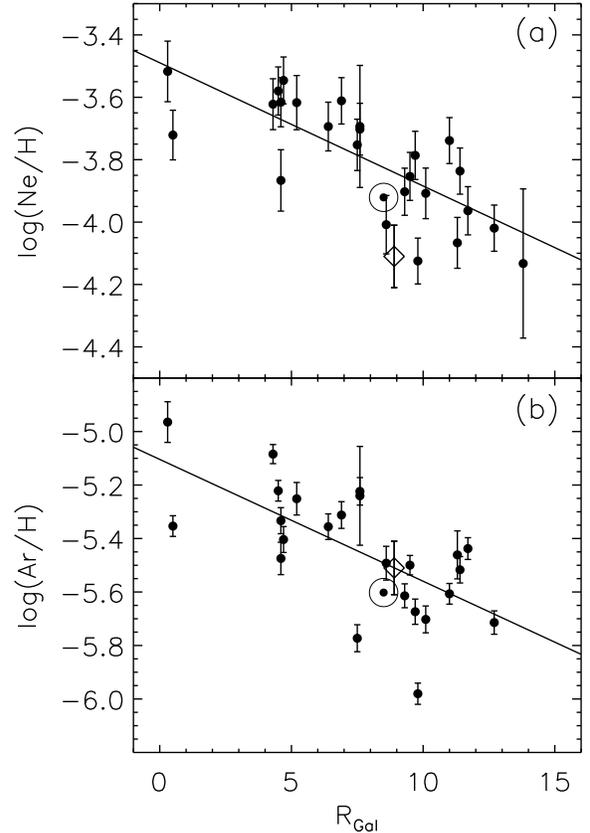,width=8.5cm}}
          \end{center}
          \vspace{-0.8cm}
          \caption{Total abundances of (a) neon and (b) argon as a
            function of \rgal. The solar
            abundance \citep{grevesse98} is indicated by the
            symbol \sun\ and the Orion abundance \citep{esteban98}
            by an open diamond.}
          \label{fig:alpha}
        \end{figure}

Fig.~\ref{fig:alpha} shows Ne/H and Ar/H as a function of \rgal. Clear
gradients with galactocentric distance can be observed for both
species.   Linear least squares fits to the neon and  argon abundances,
displayed as solid lines in  Fig. \ref{fig:alpha}, are:

\begin{equation}
{\rm log {Ne \over H} } = (-3.49 \pm 0.06) - (0.039 \pm 0.007){\rm R_{Gal}}~~ 
{\rm and}
\end{equation}

\begin{equation}
{\rm log {Ar \over H} } = (-5.10 \pm 0.09) - (0.045 \pm 0.011){\rm R_{Gal}}~,
\end{equation}

\noindent
with correlation coefficients r=$-0.76$ and $-0.66$, respectively. 

        \begin{figure}[!ht]
          \vspace{-0.55cm}
          \begin{center}
            \centerline{\psfig{file=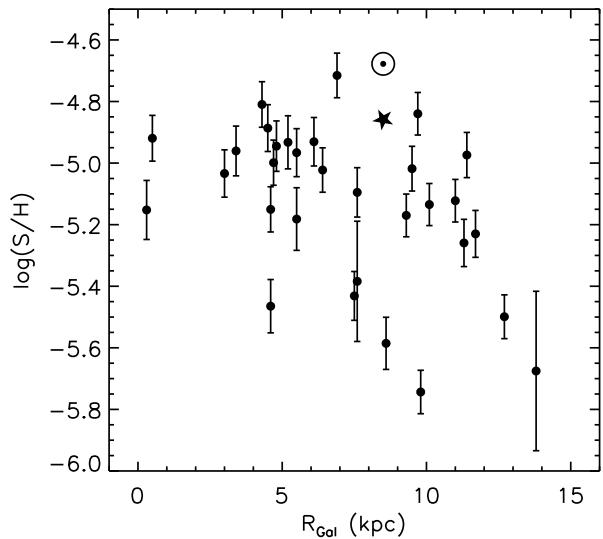,width=8.5cm}}
          \end{center}
          \vspace{-1.5cm}
          \caption{S/H abundance as a function of \rgal. The solar
            abundance \citep{grevesse98} is indicated by the
            symbol \sun\ and the interstellar abundance \citep{snow96} by a 
            star. As discussed in Sect.~\ref{Alpha}, 
            S/H could be underestimated by up to a factor 4 due to the 
            \Tel\ and \nel\ uncertainties.}
          \label{fig:sulphurab}
        \end{figure}

We note that an electron temperature gradient in the Galaxy may
systematically enlarge the above derived elemental abundance
gradients.  Specifically, if we correct for a \Tel\ gradient given by
\Tel$=5000+5000\times$(\rgal(kpc)/15), based on the \Tel\ gradients
derived by \cite{shaver83}, \cite{afflerbach96} and \cite{deharveng00}, 
the slopes in the Ne/H and Ar/H gradients
become $-0.06$ and $-0.07$ dex kpc$^{-1}$, respectively.

While Ar/H and Ne/H are \Tel-dependent through the Br$\alpha$ line
emissivity, the ratio Ar/Ne is practically insensitive to changes in
\Tel. An approximately constant Ar/Ne close to solar is found,
contrary to the tentative Ar/Ne galactic gradient derived from a
small sample of KAO observations \citep{simpson95b}.

S/H abundances are plotted in Fig.~\ref{fig:sulphurab}.  No meaningful
Galactic gradient is seen for S/H and a  very large dispersion is
found at any given \rgal, unlike the smaller dispersion in the Ar/H or
Ne/H Galactic abundances. The absolute abundances are also improbably
low compared to the solar  and interstellar abundances -- on average 2
times lower that the IS sulphur abundance by \cite{snow96}; a
factor of 5 for the extreme cases --,  since sulphur is not expected to
be strongly depleted into grains.  However, the quoted sulphur
abundances do not include the  systematic uncertainties due to the
effect of electron temperature and density
(cf. Sect.~\ref{Methodology}). In view of the rms densities  derived from
the radio observations (we note that the sources with the lower S/H
abundances also have very high \rmsnel$> 10^4$~\cm3) and the \Tel\
gradient in the Galaxy, we  estimate that the S abundances are
underestimated by up to a factor 4.

\subsection{Comparison of gradients with previous studies}
\label{Comparison}

\begin{figure}[!ht]
          \vspace{-0.55cm}
          \begin{center}
            \centerline{\psfig{file=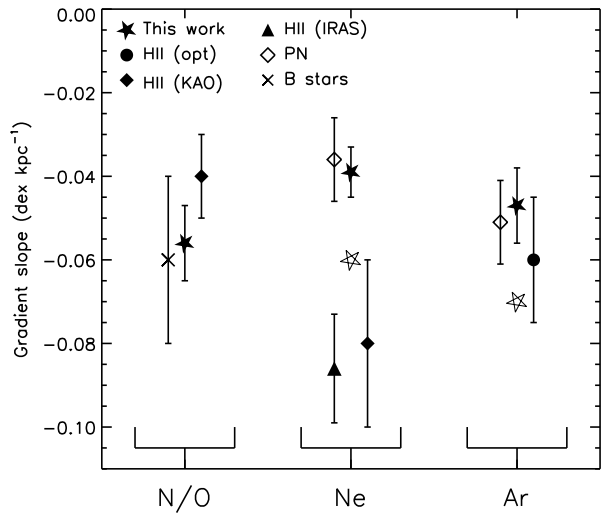,width=8.5cm}}
          \end{center}
          \vspace{-1.5cm}
          \caption{Comparison of the derived gradient slopes for N/O,
            Ne/H and Ar/H with previous determinations using
            \HII\ regions -- optical \citep{shaver83}, IRAS \citep{simpson90}, 
            KAO \citep{simpson95a} --, disk planetary nebulae of type II 
            \citep{maciel99} and B stars \citep{smartt01}.
            The open stars indicate the slopes obtained 
            when a Galactic \Tel\ gradient of $\sim$ 330 K\,kpc$^{-1}$
            is assumed.}
          \label{fig:comparison}
        \end{figure}

Fig.~\ref{fig:comparison} compares the gradient slopes derived  for
N/O, Ne/H and Ar/H with previous determinations  using \HII\ regions --
optical \citep{shaver83}, IRAS \citep{simpson90}, KAO \citep{simpson95a} --, 
disk planetary nebulae of type II \citep{maciel99} and B stars 
\citep{smartt01}.

Good agreement is found for N/O and Ar/H. Our Ne/H gradient matches
the one found from planetary nebula, but is flatter than the previous
infrared determinations using IRAS and KAO.  Both IRAS and KAO
determinations agree, however. We note that both relied on radio data
to derive the abundances and therefore, did not suffer much from the
systematic electron temperature uncertainty, while our gradients have
been probably depressed because of the assumption of a single
\Tel=7500 K. Indeed, if we correct  our derived Ne/H gradient from the
\Tel\ gradient in the Galaxy (the open stars in
Fig.~\ref{fig:comparison}), then we are in reasonable agreement with
both IRAS and KAO.

\begin{table*}[!ht]
\caption{Ionic and total abundances of neon, sulphur and argon derived
  in the low density limit and assuming a single \Tel=7500~K. 
  Systematic uncertainties may affect these abundances  because of the \Tel\
  Galactic gradient. Sulphur abundances for some sources could
  be underestimated by up to a factor 4 due to the \Tel\ and
  \nel\ uncertainties (see Sect.~\ref{Alpha}).}
\label{table:alpha}
\begin{center}
\leavevmode
\scriptsize
\begin{tabular}{lr@{\,}c@{\,}lr@{\,}c@{\,}lr@{\,}c@{\,}lr@{\,}c@{\,}lr@{\,}c@{\,}lr@{\,}c@{\,}lr@{\,}c@{\,}lr@{\,}c@{\,}lr@{\,}c@{\,}l} \hline \\
\multicolumn{1}{c}{}
&\multicolumn{9}{c}{NEON}
&\multicolumn{9}{c}{SULPHUR}
&\multicolumn{9}{c}{ARGON}\\ \\[1pt]
\multicolumn{1}{c}{Source}
& \multicolumn{3}{c}{Ne$^{+}$/H$^{+}$}
& \multicolumn{3}{c}{Ne$^{++}$/H$^{+}$ }
& \multicolumn{3}{c}{Ne/H}
& \multicolumn{3}{c}{S$^{++}$/H$^{+}$}
& \multicolumn{3}{c}{S$^{+3}$/H$^{+}$}
& \multicolumn{3}{c}{S/H}
& \multicolumn{3}{c}{Ar$^{+}$/H$^{+}$}
& \multicolumn{3}{c}{Ar$^{++}$/H$^{+}$}
& \multicolumn{3}{c}{Ar/H}\\
\multicolumn{1}{c}{ }
& \multicolumn{3}{c}{$\times 10^{-4}$}
& \multicolumn{3}{c}{$\times 10^{-4}$}
& \multicolumn{3}{c}{$\times 10^{-4}$}
& \multicolumn{3}{c}{$\times 10^{-6}$}
& \multicolumn{3}{c}{$\times 10^{-6}$}
& \multicolumn{3}{c}{$\times 10^{-6}$}
& \multicolumn{3}{c}{$\times 10^{-6}$}
& \multicolumn{3}{c}{$\times 10^{-6}$}
& \multicolumn{3}{c}{$\times 10^{-6}$} \\[2pt] \hline \\[1pt]
IR\,01045 &   0.6 &  $\pm$ &   0.3 &   0.16 &  $\pm$ &   0.09 &   0.7 & 
 $\pm$ &   0.4 &   2. &  $\pm$ &   1. &   &  $<$ &  0.1 &   2. &  $\pm$ &   1.
 &   &  $<$ &  0.6 &   &  $<$ &  0.5 &   &  $<$ &  1.\\
IR\,02219 &   0.9 &  $\pm$ &   0.2 &   0.9 &  $\pm$ &   0.2 &   1.8 & 
 $\pm$ &   0.3 &   7. &  $\pm$ &   1. &   0.90 &  $\pm$ &   0.09 &   8. & 
 $\pm$ &   1. &   0.36 &  $\pm$ &   0.03 &   2.1 &  $\pm$ &   0.2 &   2.5 & 
 $\pm$ &   0.2\\
IR\,02575 &   &  $<$ &  0.1 &   &  $<$ &  0.04 &   &  $<$ &  0.1 &   & 
 $<$ &  0.5 &   &  $<$ &  0.09 &   &  $<$ &  0.6 &   &  $<$ &  0.5 &   & 
 $<$ &  0.3 &   &  $<$ &  0.8\\
IR\,10589 &   1.2 &  $\pm$ &   0.2 &   0.16 &  $\pm$ &   0.03 &   1.4 & 
 $\pm$ &   0.2 &   9. &  $\pm$ &   2. &   0.18 &  $\pm$ &   0.03 &  10. & 
 $\pm$ &   2. &   1.7 &  $\pm$ &   0.1 &   1.5 &  $\pm$ &   0.2 &   3.2 & 
 $\pm$ &   0.3\\
IR\,11143 &   0.54 &  $\pm$ &   0.10 &   1.1 &  $\pm$ &   0.2 &   1.6 & 
 $\pm$ &   0.3 &  12. &  $\pm$ &   2. &   2.6 &  $\pm$ &   0.3 &  14. & 
 $\pm$ &   2. &   &  $<$ &  0.2 &   2.1 &  $\pm$ &   0.2 &   2.1 & 
 $\pm$ &   0.2\\
IR\,12063 &   0.7 &  $\pm$ &   0.1 &   0.59 &  $\pm$ &   0.10 &   1.3 & 
 $\pm$ &   0.2 &   5.8 &  $\pm$ &   1.0 &   0.92 &  $\pm$ &   0.09 &   7. & 
 $\pm$ &   1. &   0.43 &  $\pm$ &   0.05 &   2.0 &  $\pm$ &   0.2 &   2.4 & 
 $\pm$ &   0.3\\
IR\,12073 &   0.34 &  $\pm$ &   0.06 &   0.9 &  $\pm$ &   0.2 &   1.2 & 
 $\pm$ &   0.2 &   5.4 &  $\pm$ &   1.0 &   1.9 &  $\pm$ &   0.2 &   7. & 
 $\pm$ &   1. &   0.12 &  $\pm$ &   0.04 &   1.9 &  $\pm$ &   0.2 &   2.0 & 
 $\pm$ &   0.2\\
IR\,12331 &   1.9 &  $\pm$ &   0.3 &   0.59 &  $\pm$ &   0.10 &   2.4 & 
 $\pm$ &   0.4 &  19. &  $\pm$ &   3. &   0.53 &  $\pm$ &   0.05 &  19. & 
 $\pm$ &   3. &   1.6 &  $\pm$ &   0.2 &   3.3 &  $\pm$ &   0.4 &   4.9 & 
 $\pm$ &   0.6\\
IR\,15384 &   1.8 &  $\pm$ &   0.3 &   0.25 &  $\pm$ &   0.05 &   2.0 & 
 $\pm$ &   0.4 &   9. &  $\pm$ &   2. &   0.19 &  $\pm$ &   0.02 &   9. & 
 $\pm$ &   2. &   2.6 &  $\pm$ &   0.2 &   1.8 &  $\pm$ &   0.3 &   4.4 & 
 $\pm$ &   0.5\\
IR\,15502 &   1.2 &  $\pm$ &   0.3 &   0.13 &  $\pm$ &   0.03 &   1.4 & 
 $\pm$ &   0.3 &   3.4 &  $\pm$ &   0.7 &   0.033 &  $\pm$ &   0.006 &   3.4
 &  $\pm$ &   0.7 &   2.7 &  $\pm$ &   0.4 &   0.6 &  $\pm$ &   0.1 &   3.4 & 
 $\pm$ &   0.5\\
IR\,17221 &   2.4 &  $\pm$ &   0.5 &   0.04 &  $\pm$ &   0.01 &   2.4 & 
 $\pm$ &   0.5 &  12. &  $\pm$ &   2. &   0.044 &  $\pm$ &   0.010 &  12. & 
 $\pm$ &   2. &   4.4 &  $\pm$ &   0.6 &   1.2 &  $\pm$ &   0.2 &   5.6 & 
 $\pm$ &   0.8\\
 Sgr\,C &   3.0 &  $\pm$ &   0.7 &   &  $<$ &  0.05 &   3.0 &  $\pm$ &   0.7
 &   7. &  $\pm$ &   2. &   &  $<$ &  0.08 &   7. &  $\pm$ &   2. &  11. & 
 $\pm$ &   2. &   &  $<$ &  0.4 &  11. &  $\pm$ &   2.\\
IR\,17455 &   1.6 &  $\pm$ &   0.3 &   0.31 &  $\pm$ &   0.05 &   1.9 & 
 $\pm$ &   0.3 &  12. &  $\pm$ &   2. &   0.19 &  $\pm$ &   0.03 &  12. & 
 $\pm$ &   2. &   2.7 &  $\pm$ &   0.2 &   1.8 &  $\pm$ &   0.2 &   4.4 & 
 $\pm$ &   0.4\\
IR\,18032 &   2.0 &  $\pm$ &   0.4 &   &  $<$ &  0.02 &   2.0 &  $\pm$ &   0.4
 &   8. &  $\pm$ &   1. &   &  $<$ &  0.03 &   8. &  $\pm$ &   1. &   5.3 & 
 $\pm$ &   0.6 &   0.71 &  $\pm$ &   0.09 &   6.0 &  $\pm$ &   0.7\\
IR\,18116 &   2.3 &  $\pm$ &   0.4 &   0.08 &  $\pm$ &   0.02 &   2.4 & 
 $\pm$ &   0.4 &  15. &  $\pm$ &   3. &   &  $<$ &  0.03 &  15. & 
 $\pm$ &   3. &   7.4 &  $\pm$ &   0.6 &   0.9 &  $\pm$ &   0.1 &   8.2 & 
 $\pm$ &   0.7\\
IR\,18317 &   2.6 &  $\pm$ &   0.5 &   0.08 &  $\pm$ &   0.01 &   2.6 & 
 $\pm$ &   0.5 &  13. &  $\pm$ &   2. &   0.034 &  $\pm$ &   0.009 &  13. & 
 $\pm$ &   2. &   4.9 &  $\pm$ &   0.4 &   1.1 &  $\pm$ &   0.1 &   6.0 & 
 $\pm$ &   0.5\\
IR\,18434 &   2.2 &  $\pm$ &   0.4 &   0.26 &  $\pm$ &   0.05 &   2.4 & 
 $\pm$ &   0.4 &   7. &  $\pm$ &   1. &   0.13 &  $\pm$ &   0.02 &   7. & 
 $\pm$ &   1. &   3.0 &  $\pm$ &   0.3 &   1.7 &  $\pm$ &   0.2 &   4.6 & 
 $\pm$ &   0.5\\
IR\,18479 &   1.5 &  $\pm$ &   0.3 &   0.30 &  $\pm$ &   0.06 &   1.8 & 
 $\pm$ &   0.3 &   3.7 &  $\pm$ &   0.7 &   &  $<$ &  0.05 &   3.7 & 
 $\pm$ &   0.7 &   1.7 &  $\pm$ &   0.2 &   &  $<$ &  0.2 &   1.7 & 
 $\pm$ &   0.2\\
IR\,18502 &   2.6 &  $\pm$ &   0.4 &   0.26 &  $\pm$ &   0.05 &   2.8 & 
 $\pm$ &   0.5 &  10. &  $\pm$ &   2. &   0.12 &  $\pm$ &   0.02 &  10. & 
 $\pm$ &   2. &   2.9 &  $\pm$ &   0.3 &   1.1 &  $\pm$ &   0.1 &   3.9 & 
 $\pm$ &   0.4\\
IR\,19442 &   2.0 &  $\pm$ &   0.9 &   &  $<$ &  0.03 &   2.0 &  $\pm$ &   0.9
 &   4. &  $\pm$ &   2. &   &  $<$ &  0.1 &   4. &  $\pm$ &   2. &   6. & 
 $\pm$ &   2. &   &  $<$ &  0.3 &   6. &  $\pm$ &   2.\\
IR\,19598 &   0.37 &  $\pm$ &   0.06 &   0.38 &  $\pm$ &   0.06 &   0.8 & 
 $\pm$ &   0.1 &   1.7 &  $\pm$ &   0.3 &   0.15 &  $\pm$ &   0.01 &   1.8 & 
 $\pm$ &   0.3 &   0.43 &  $\pm$ &   0.03 &   0.61 &  $\pm$ &   0.06 &   1.05
 &  $\pm$ &   0.10\\
 DR\,21 &   0.9 &  $\pm$ &   0.2 &   0.11 &  $\pm$ &   0.05 &   1.0 & 
 $\pm$ &   0.2 &   2.6 &  $\pm$ &   0.5 &   &  $<$ &  0.02 &   2.6 & 
 $\pm$ &   0.5 &   2.9 &  $\pm$ &   0.4 &   0.35 &  $\pm$ &   0.07 &   3.2 & 
 $\pm$ &   0.5\\
IR\,21190 &   0.7 &  $\pm$ &   0.1 &   0.29 &  $\pm$ &   0.05 &   1.0 & 
 $\pm$ &   0.2 &   3.0 &  $\pm$ &   0.5 &   0.20 &  $\pm$ &   0.02 &   3.2 & 
 $\pm$ &   0.5 &   0.50 &  $\pm$ &   0.06 &   1.4 &  $\pm$ &   0.1 &   1.9 & 
 $\pm$ &   0.2\\
IR\,22308 &   0.9 &  $\pm$ &   0.2 &   &  $<$ &  0.02 &   0.9 &  $\pm$ &   0.2
 &   5.5 &  $\pm$ &   1.0 &   &  $<$ &  0.05 &   5.5 &  $\pm$ &   1.0 &   2.8
 &  $\pm$ &   0.5 &   0.6 &  $\pm$ &   0.2 &   3.5 &  $\pm$ &   0.7\\
IR\,23030 &   1.4 &  $\pm$ &   0.2 &   0.09 &  $\pm$ &   0.01 &   1.5 & 
 $\pm$ &   0.2 &  10. &  $\pm$ &   2. &   0.17 &  $\pm$ &   0.02 &  11. & 
 $\pm$ &   2. &   1.0 &  $\pm$ &   0.1 &   2.0 &  $\pm$ &   0.2 &   3.0 & 
 $\pm$ &   0.3\\
IR\,23133 &   1.1 &  $\pm$ &   0.2 &   &  $<$ &  0.01 &   1.1 &  $\pm$ &   0.2
 &   6. &  $\pm$ &   1. &   0.025 &  $\pm$ &   0.006 &   6. &  $\pm$ &   1. & 
  2.9 &  $\pm$ &   0.3 &   0.73 &  $\pm$ &   0.08 &   3.7 &  $\pm$ &   0.3\\
\\ \hline
\end{tabular}
\end{center}
\end{table*}


\section{Summary and conclusions}
\label{Conclusions}

Based on the ISO spectral catalogue of compact \HII\ regions by
\cite{peeters:catalogue}, a first analysis of the hydrogen recombination
and atomic fine-structure lines originated in the ionized gas has
been presented. The main results of this study are:

\vspace{-0.5cm}
\paragraph{$\bullet$}

The SWS \HI\ recombination lines between 2 and 8 \mum\ have been used
to estimate the extinction law at these wavelengths for 14 \HII\
regions. An extinction in the K band between 0 and $\sim$ 3 mag. has been
derived.

\vspace{-0.5cm}
\paragraph{$\bullet$}

The electron densities determined by the \OIII\ 88/52~\mum\ line
ratio, in between $\sim 100$ and 3000 \cm3,  are shown to be
substantially smaller than the rms densities resulting from radio
continuum observations. This fact suggests that the ultracompact cores
observed in the radio have a region of less dense gas (traced by
\OIII\ 88/52 \mum) at high excitation, which could be produced by
leakage of radiation from the ultracompact region.   This low density
gas might be identified with the physically related, low brightness
envelopes recently mapped in radio around ultracompact \HII\ regions
\citep{kurtz99,kim01}.

\vspace{-0.5cm}
\paragraph{$\bullet$}

Significant variations are found in the degree of ionization of
the \HII\ regions as measured by the ratios \arionic, \nionic, \sionic\
and \neionic. These ratios, which span a range of ionization potential up
to 41~eV, correlate well with each other, suggesting that
the spectral hardening affects equally the full range of
ionizing energies. In addition, the ionic ratios vary with \rgal,
with the \HII\ regions in the outer Galaxy being powered by a
seemingly harder radiation field.
This increase in degree of ionization, ie.
in properties of the ionizing star, 
could be related to the metallicity gradient.
However, the precise mechanism
coupling the stellar characteristics to the elemental abundances
remains unclear.  Elemental abundances may couple directly to the
stellar atmosphere opacity which controls the flux and its dependence
on the frequency.  Qualitatively, theoretical models support this
coupling \citep[e.g.][]{balick76}.  The opacity by trace ions
also controls the stellar wind, which in their turn have a major
influence on the emerging stellar flux.   Again, models show
qualitative agreement \citep[e.g.][]{schaerer97}.  Third, the
connection between elemental abundances and stellar flux may be
indirect.  For example, the elemental abundance may control the dust
opacity and thereby the mass of the star formed because of radiation
pressure effects during the collapse phase \citep[cf.][]{kahn74,shields76}.  
Last, the hardening of the radiation field with \rgal\
may reflect a systematic variation in the ``age'' of the star powering
the \HII\ regions.  Possibly, in the inner parts of the Galaxy
ionizing stars evolve more ``rapidly'' (ie. are cooler for the same
luminosity) than in the outer Galaxy.  In that respect, the  detailed
photoionized model for the \HII\ region  IRAS\,18434$-$0242
(G29.96$-$0.02), using recent line-blanketed, non-LTE stellar
atmosphere models including winds \citep{schaerer97},  finds
good agreement with all the ion pair ratios using a luminous star
which has already evolved off the main sequence (Paper~III).  In such
a model, the variations in the degree of ionization with \rgal\ might
reflect a selection effect in our sample where, for example, the rate
of expansion of the compact \HII\ region is linked to the environment
(ie. metallicity).

\vspace{-0.5cm}
\paragraph{$\bullet$}

A gradient in \noionic\ is observed in the sense of a decreasing ionic
abundance ratio with \rgal.  This observed ionic abundance ratio
\noionic\ has been converted into the elemental abundance ratio N/O
using the observed ionization ratio  \neionic, sensitive to the
ionization state of the \HII\ region,  and detailed photoionization
model calculations.  The N/O abundance ratio also shows a decrease
with \rgal, although shallower than in the case of \noionic, pointing
to the existence of an  intrinsic variation in the gas phase N/O
abundance.  Thus, there seems to be an important contribution to the
synthesis of nitrogen by secondary production.  The derived slope for
N/O ($\Delta \rm log \, N/O = - 0.056\pm 0.009 \, dex \,kpc^{-1}$)
is in agreement with previous \HII\ regions \citep[e.g.][]{simpson95a} 
and B stars \citep[e.g.][]{smartt01} determinations.

\vspace{-0.5cm}
\paragraph{$\bullet$}

Our analysis method underestimated the sulphur abundances by up to a
factor 4 and hence, no meaningful Galactic gradient could be
derived. We estimated  a lower limit for the oxygen abundance through
the derivation of O$^{++}$/H$^+$. We miss the important  \ion{O}{ii}
ionization stage, which does not have any infrared fine-structure
line.  On the other hand, because the \NII\ 122~\mum\ line is very
weak and it was only detected in a limited number of sources,  and no
radio information was available for all these sources,  N/H could only
be determined for a few sources and therefore, a gradient could not be
drawn. We found, however, supersolar nitrogen abundances for the
sources in the inner 5 kpc.

\vspace{-0.5cm}
\paragraph{$\bullet$}

The total elemental abundances of neon and argon with respect to H
have been also derived. Ne/H and Ar/H have been found to decrease with
the galactocentric distance as  $\Delta \rm log \, Ne/H=- 0.039\pm
0.007 \, dex \,kpc^{-1}$ and  $\Delta \rm log \, Ar/H=-0.045\pm 0.011
\, dex \,kpc^{-1}$.  The slopes of neon and argon coincide within the
uncertainties, indicating that both species have a similar stellar
nucleosynthesis history.  However, the existence of a Galactocentric
\Tel\ gradient may enlarge  these elemental gradients.  Because only a
few radio recombination line observations towards the \HII\ regions,
which can be used to characterize the nebular \Tel,  are available, we
do not have enough information to correct individually the derived
abundances.   However, adopting a decrease in \Tel\ of $\sim$ 330 K
kpc$^{-1}$ with galactocentric distance, we estimate that the  slopes
for the Ne/H and Ar/H gradients become $-0.06$ and  $-0.07$
dex\,kpc$^{-1}$, respectively. These corrected slopes are in
reasonable agreement with the previous determinations by  IRAS
\citep{simpson90} and KAO \citep{simpson95a}. Ar/Ne, which
is independent of \Tel, is found to be approximately solar and constant with
\rgal.

\vspace{-0.5cm}
\paragraph{}

The complete infrared spectra from 2.3 to 196~\mum\ provided by the
ISO catalogue of galactic \HII\ regions \citep{peeters:catalogue} 
has allowed us to study the ionizing conditions in these
sources and derive relative and absolute elemental abundances across
the Galactic disk. The first analysis presented in this paper has
shown that useful conclusions can be drawn from these data. However,
there are still open issues and to improve on the present results
additional information is required. In particular, the detailed
knowledge of the extent and relative spatial distributions of the
fine-structure atomic lines is lacking for almost all the compact
\HII\ regions.  
The knowledge of the electron temperature is also
critical in refining the elemental abundances. Observations of radio
recombinations lines will be helpful to estimate \Tel\ for all the
sources discussed in this paper and to correct for the \Tel\
dependence.

\end{document}